\documentclass[reprint,10pt,nofootinbib,notitlepage,superscriptaddress,aps,prd]{revtex4-2}
\usepackage{amsmath}
\usepackage{amsfonts}
\usepackage{amssymb}
\usepackage[utf8]{inputenc}
\usepackage[T1]{fontenc}
\usepackage{xcolor}
\usepackage[colorlinks=true, final=true, pdfnewwindow=true]{hyperref}
\usepackage{url}
\usepackage{enumerate}
\usepackage{enumitem}
\usepackage{graphicx}
\usepackage{acronym}
\usepackage{subfigure}
\usepackage{cleveref}

\usepackage{ragged2e}

\newcommand{\Emax}{E_\text{max}}
\newcommand{\Rmax}{R_\text{max}}
\newcommand{\change}[1]{\textcolor{black}{#1}}
\newcommand{\mchange}{\color{black}}

\begin{document}

\title{ 
\change{Effects of extragalactic magnetic field on the spectra of ultra-high-energy cosmic rays from jetted
sources}}

\author{Sarah Soares Sippert}
\email{sarahsippert@pos.if.ufrj.br}
\affiliation{Instituto de Física, Universidade Federal do Rio de Janeiro, Rio de Janeiro, Rio de Janeiro, Brazil}

\author{Carlos Magno R. da Costa}
\email{magno_costa@id.uff.br}
\affiliation{Instituto de Ciências Exatas, Universidade Federal Fluminense, Volta Redonda, Rio de Janeiro, Brazil}

\author{Rogerio M. de Almeida}
\email{rmenezes@if.ufrj.br}
\affiliation{Instituto de Física, Universidade Federal do Rio de Janeiro, Rio de Janeiro, Rio de Janeiro, Brazil}

\author{Rafael \surname{Alves Batista}}
\email{rafael.alves\_batista@iap.fr}
\affiliation{Sorbonne Université, CNRS, UMR 7095, Institut d’Astrophysique de Paris (IAP), Paris, France }
\affiliation{Sorbonne Université, Laboratoire de Physique Nucléaire et de Hautes Energies (LPNHE), CNRS, Paris, France}

\author{João R. T. de Mello Neto}
\email{jtmn@if.ufrj.br}
\affiliation{Instituto de Física, Universidade Federal do Rio de Janeiro, Rio de Janeiro, Rio de Janeiro, Brazil}

\date{\today}

\begin{abstract}

The origins and acceleration mechanisms of ultra-high-energy cosmic rays (UHECRs) are unknown. Many models attribute their extreme energies to powerful astrophysical jets. Understanding whether jet geometry -- specifically the opening angle and its orientation relative to Earth -- affects observational signatures is crucial for interpreting UHECR data.
In this work, we perform numerical simulations of UHECR propagation in a magnetized universe to investigate the spectral signatures of jetted and nonjetted astrophysical sources. We demonstrate, for the first time, that under certain conditions, emission geometry can play a decisive role in shaping the observed spectrum of individual UHECR sources. 
These findings provide new insights into the conditions necessary for detecting UHECRs from jets, and highlight how the interplay between emission geometry and magnetic fields influences observed energy spectra.

\end{abstract}

\maketitle

\acrodef{AGN}{active galactic nucleus}
\acrodef{CL}{confidence level}
\acrodef{CMB}{cosmic microwave background}
\acrodef{CR}{cosmic ray}
\acrodef{CRB}{cosmic radio background}
\acrodef{EAS}{extensive air shower}
\acrodef{EBL}{extragalactic background light}
\acrodef{EGMF}{extragalactic magnetic field}
\acrodef{GCR}{Galactic cosmic ray}
\acrodef{GMF}{Galactic magnetic field}
\acrodef{GRB}{gamma-ray burst}
\acrodef{MHD}{magnetohydrodynamics}
\acrodef{GZK}{Greisen-Zatsepin-Kuzmin}
\acrodef{HE}{high energy}
\acrodef{RMS}{root mean square}
\acrodef{SBG}{starburst galaxy}
\acrodef{SN}{supernova}
\acrodef{SNR}{supernova remnant}
\acrodef{TA}{Telescope Array}
\acrodef{UHE}{ultra-high energy}
\acrodef{UHECR}{ultra-high-energy cosmic ray}
\acrodef{VHE}{very-high energy}

\section{Introduction}\label{sec:intro}

\Acp{UHECR} are particles with energies above $1 \; \text{EeV}$\footnote{1~EeV $\equiv$ $10^{18} \; \text{eV}$.} that hit our atmosphere coming from space. Since their discovery in 1963~\cite{linsley1963evidence}, their origins and the mechanisms capable of accelerating them to such extraordinary energies remain unknown, posing one of the most challenging mysteries in modern astrophysics~\cite{kotera2011astrophysics, alves2019open}.

During their journey from their accelerating sites to Earth, \acp{CR} experience adiabatic energy losses due to the cosmological expansion of the universe, and they undergo interactions with pervasive radiation, predominantly from the \ac{CMB} and the \ac{EBL}. In the case of protons, energy losses occur primarily through interactions with the \ac{CMB}, particularly from photopion production at the highest energies, and Bethe-Heitler pair production down to EeV~energies. For nuclei, the dominant energy-loss process is photodisintegration resulting from collisions with \ac{EBL} and \ac{CMB} photons. Additionally, \acp{CR} may be deflected by magnetic fields in intergalactic space, leading to a loss of directional information by the time they reach Earth, which complicates the task of identifying their origins.

The Pierre Auger Observatory~\cite{aab2015pierre} and the \ac{TA}~\cite{abu2012surface} have made significant contributions to advancing our understanding of \acp{UHECR}. In particular, with respect to searches for anisotropies in the \ac{CR} arrival direction, the Pierre Auger Observatory has reported a $6.8 \sigma$ confidence level dipolar anisotropy in the cosmic-ray flux of events detected with energies above 8~EeV. The measured dipole direction points \change{away from} the \change{Galactic Center} $(\alpha = (100 \pm 10)^\circ$, $\delta = (-24 \pm 12)^\circ)$, which is a strong evidence for the dominance of extragalactic sources in the \ac{UHECR} flux above this energy~\cite{halim2024large}. Besides, in the context of intermediate scale anisotropies,  the Auger Collaboration reported a correlation between \acp{UHECR} with energies above 39~EeV and active galactic nuclei, with a 3.3$\sigma$ confidence level, and an even stronger 4.2$\sigma$ correlation with \acp{SBG}~\cite{szadkowski2022arrival}. Furthermore, the correlation with \acp{SBG} increases to $4.4\sigma$ if a full-sky search is performed by combining Auger and \ac{TA} data~\cite{rubtsov2025update}. Despite these significant advances, the sources of \acp{UHECR} remain elusive and their precise origins have yet to be conclusively identified.

The mechanisms behind the acceleration of \acp{UHECR} are still being investigated. Among the candidate sources of these particles are the astrophysical jets. These jets are often supersonic, so they can produce strong shock waves. They create environments where velocity shear and turbulence are important and also transport strong magnetic fields. Each of these effects can lead to the acceleration of high-energy particles. On this matter, many studies have been attempted in order to understand the precise mechanisms by which particles are accelerated within these jets. They suggest that these particles could be accelerated through various processes, each operating in a different region. Magnetic reconnection at the highly magnetized jet base~\cite{giannios2010uhecrs}, diffusive shock acceleration in regions such as the jet beam, backflow, and termination shocks~\cite{ostrowski1998acceleration}, and shear acceleration in high-speed gradient regions~\cite{rieger2004shear} are possible contributors.

The acceleration processes and how cosmic rays interact with the magnetic fields in these environments remain complex.  As \acp{UHECR} travel through the jet's turbulent lobes, they may lose their directional information through scattering, but this depends on the scattering length and the coherence of the magnetic field. Although the interaction with turbulent magnetic fields and scattering in the lobes tends to isotropize the emission, certain acceleration mechanisms can cause the cosmic rays to retain a directional preference. For instance, \acp{UHECR} accelerated in relativistic blobs within the jet are expected to experience relativistic beaming, potentially resulting in anisotropic emission in the laboratory frame, depending on the observer’s viewpoint ~\cite{orr1982relativistic}. This nonisotropic emission could offer crucial insights into both the acceleration mechanisms and the jet’s underlying geometry.

In this work, assuming that \acp{UHECR} are accelerated and emitted along the jet direction, we investigate the effect of the geometry of cosmic-ray emission and the extragalactic magnetic field properties along the propagation on the observed energy spectrum. In Sec.~\ref{sec:diffusion}, we briefly discuss \ac{CR} diffusion in the context of a particle propagation in a turbulent magnetic field. Section~\ref{sec:simulation} describes the simulations used in this work. The results are presented and discussed in Sec.~\ref{sec:results}. Finally, the conclusions are presented in Sec.~\ref{sec:conclusions}.

\section{Diffusive Cosmic Ray Propagation in Turbulent Magnetic Fields}
\label{sec:diffusion}

The probability of a particle reaching Earth ultimately depends on the properties of the magnetic field along its path. Therefore, understanding the propagation of \acp{UHECR} through \acp{EGMF} is essential for a correct interpretation of the results presented in this work. However, the precise characteristics of these fields are still uncertain due to limited observational data. 

Magnetic fields in galaxy clusters range from a few to tens of~$\mu$G in their cores, down to  $1-10$~nG in the outskirts~\cite{carilli2002cluster, vacca2012intracluster, guidetti2008intracluster, donnert2018a}. Estimates of magnetic fields in the filaments connecting galaxy clusters point to strengths of approximately $30-100$~nG~\cite{govoni2019radio, vernstrom2021a, carretti2022a}. Recent observational studies on magnetic fields in less dense regions, such as cosmic voids, estimate strengths ranging from $10^{-5}$ to $10^{-3}$~pG~\cite{hosking2023cosmic}, although \ac{CMB} constraints place an upper bound of $\sim 10 \; \text{pG}$~\cite{jedamzik2019a}. \change{In addition, upper bounds on the \ac{EGMF} can also be derived from rotation measures of extragalactic sources~\cite{pshirkov2016new}, which probe magnetic fields at $z \sim 1$.} The coherence length of void magnetic fields can vary enormously, with only weak constraints existing, mildly favoring values between 100~kpc and 100~Mpc~\cite{alvesbatista2020a}.
In the Local Supercluster, estimates suggest that the \ac{RMS} strength of magnetic fields lies between 1 and 100~nG~\cite{vallee2002a, vallee2011a}, with coherence lengths ranging from 10~kpc to 1~Mpc.

\change{The \ac{GMF} is relatively strong compared to the extragalactic fields, reaching strengths of a few $\mu$G~\cite{jansson2012a, jansson2012b, unger2024b, korochkin2025a}. However, it does not significantly affect the \ac{CR} spectrum measured on Earth.}

Due to the limited knowledge and significant uncertainties regarding the structure and properties of extragalactic magnetic fields of large-scale environments, such as cosmic filaments, we simplify our study by focusing on the motion of \acp{CR} through a turbulent extragalactic magnetic field. This simplification assumes that the field can be fully characterized by its \ac{RMS} strength, $B \equiv \sqrt{\left< B^2(x)\right>}$, and its coherence length, $L_B$, which defines the maximum spatial scale over which the magnetic field remains correlated. Although we acknowledge that this may \change{not} be a fully accurate approximation, \change{it still captures the dominant features of turbulent scattering and thus provides a reasonable framework for modeling \ac{CR} propagation.} 

When a cosmic ray propagates through a turbulent \change{\ac{EGMF}}, it may undergo multiple deflections. This behavior can be treated as a random walk, allowing the definition of a diffusion coefficient and an associated diffusion length, which corresponds to the average distance a particle travels before experiencing a significant deflection, providing a framework to describe its motion. The deviation of the particle's trajectory after each deflection depends on its energy, charge, and the strength and coherence length of the magnetic field. To facilitate the analysis of diffusive cosmic-ray propagation in such environments, a characteristic energy scale, known as the critical energy ($E_\text{c}$), is introduced.

The critical energy $E_\text{c}$ is defined as the energy for which the particle's Larmor radius ($r_\text{L}$),
\begin{equation}
    r_\text{L} = \dfrac{E}{ZeB} \approx  \dfrac{1.1}{Z} \left(\dfrac{E}{\text{EeV}}\right) \left(\dfrac{B}{\text{nG}}\right)^{-1} \; \text{Mpc}  \,,
\end{equation}
equals the coherence length of the magnetic field ($L_B$), and is given by
\begin{equation}
\label{critical_energy}
E_\text{c} = Z e  B L_B \approx 0.9 Z \ \left(\dfrac{B}{\rm nG}\right) \ \left(\dfrac{L_B}{\rm Mpc}\right) \; { \rm EeV} \,.
\end{equation}

The diffusion coefficient $D(E)$ is given by $D(E) = c l_\text{D}/3$, where $l_\text{D}$ is the diffusion length. Reference~\cite{harari2014anisotropies} proposed a fit for the diffusion coefficient that closely matches the numerical integration results: 
\begin{equation}\label{eq:difusao}
    D(E) \simeq\frac{cL_{B}(z)}{3}\left[4\left(\frac{E}{E_\text{c}}\right)^{2} + a_{I}\left(\frac{E}{E_\text{c}}\right) +  a_{L}\left(\frac{E}{E_\text{c}}\right)^{\alpha}\right].
\end{equation}
Consequently the diffusion length is given by
\begin{equation}\label{eq:difusion_length}
    l_\text{D}(E) \simeq L_B \left[4\left(\frac{E}{E_\text{c}}\right)^{2} + a_{I}\left(\frac{E}{E_\text{c}}\right) +  a_{L}\left(\frac{E}{E_\text{c}}\right)^{\alpha}\right],
\end{equation}
where the coefficients $a_I$ and $a_L$ are respectively  $\approx 0.9$ and $\approx 0.23$ for a Kolmogorov spectrum. The value of exponent $\alpha$ according to Kolmogorov turbulence is $\alpha = 1/3$~\cite{Kolmogorov1941}.

Below the critical energy $E_\text{c}$, the Larmor radius is smaller than the coherence length and the particle undergoes stronger deflections. As a result, the flux from a given source becomes suppressed due to magnetic horizon effects~\cite{alvesbatista2014b}. This is quantified by the diffusion length $l_\text{D}$, in this regime $l_\text{D} \approx L_B(E_\text{c}/E)^\alpha$. For energies above $E_\text{c}$, the Larmor radius exceeds the coherence length, and the cosmic rays experience nonresonant scattering, yielding $l_\text{D} \approx L_B(E_\text{c}/E)^2$, indicating a rapid increase of diffusion length with energy. Therefore, if $l_\text{D}$ is much smaller than the distance between the source and the observer $R$, spatial diffusion occurs; if $l_\text{D} > R$, the particles propagate quasirectilinearly.

\section{Simulations}\label{sec:simulation}

We use the CRPropa framework~\cite{alvesbatista2016a, alvesbatista2022a} to simulate the propagation of UHECRs emitted from a single source, embedded in a turbulent magnetised environment. 
We take into account all relevant energy-loss mechanisms and particle interactions, i.e., Bethe-Heitler pair production, photopion production, photodisintegration, nuclear decay, and adiabatic energy losses due to the expansion of the universe. Particle propagation is terminated if the particle’s energy falls below 0.1~EeV or if the propagation time exceeds the age of the Universe.

The source is assumed to emit \acp{CR} according to an injection spectrum of the form 
\begin{equation}
    \dfrac{\text{d}N}{\text{d}E} \propto E^{-\gamma} f_{\text{cut}}(E, Z) \,,
    \label{eq:specInjection}
\end{equation}
where $\gamma$ is the spectral index of the source and $f_{\text{cut}}$ is the cutoff function, which depends on the charge ($Ze$) of the \ac{CR} nucleus. We also assume that the cosmic-ray luminosity is constant over time.

The magnetic field is assumed to have a Kolmogorov power spectrum, with coherence length $L_B$ and a \ac{RMS} field strength $B$. It is stored in a three-dimensional cubic grid of \change{10~kpc} cell spacing, within $10$~Mpc box that repeats itself periodically. We perform the simulations considering several different realizations of the magnetic field, to reduce fluctuations and to properly capture the statistical properties of the problem.

\begin{figure}[htb]
    \centering
    \includegraphics[width=0.75\columnwidth]{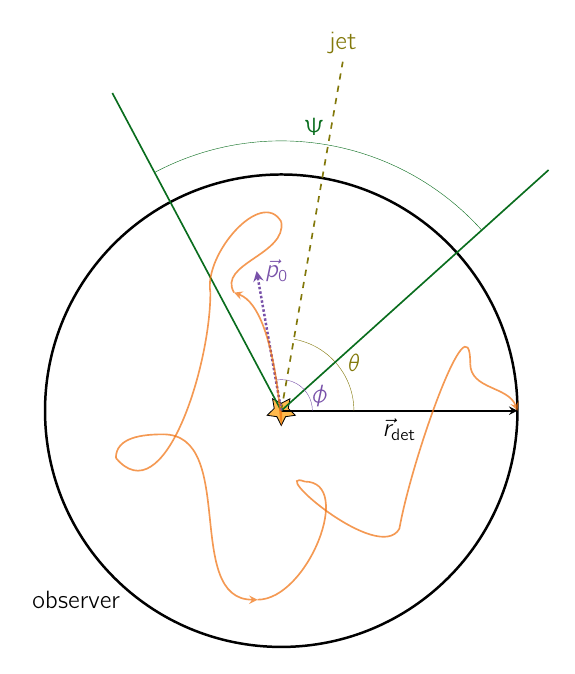}
   \caption{\justifying Illustration of the simulation setup. Particles are injected isotropically from the source (yellow star), and are collected as they cross the surface of a sphere with radius $R \equiv \left| \vec{r}_{\text{det}}\right|$. Each detection position $\vec{r}_\text{det}$ represents a possible Earth-like observer location. A simulated event with initial momentum $\vec{p}_0$ (dotted purple line) that satisfies the angular condition $\theta - \Psi/2 \leq \phi \leq \theta + \Psi/2$ is considered to originate from a jet with opening angle $\Psi$, oriented at an angle $\theta$ relative to the line of sight between the source and the Earth. The orange line illustrates the trajectory of an event.}
    \label{Fig:geom_reversa}
 \end{figure}

Since the size of our Galaxy is negligible compared to the vastness of the Universe, performing a three-dimensional forward simulation of particle propagation from a specific extragalactic location to Earth is extremely time consuming and nearly impossible to achieve. To overcome this issue and to estimate the observed energy spectrum from a given source located at a distance $R$, emitting particles in a jet with an opening angle of $\Psi$, we use the large sphere approach from Ref.~\cite{alvesbatista2016b}. First, cosmic rays are emitted isotropically from a single source. Particles are propagated and collected when they cross the surface of a sphere with radius $R$. This is done in CRPropa by defining \change{an} observer as a sphere of radius $R$ centered at the source location. Figure~\ref{Fig:geom_reversa} illustrates the geometry considered in the simulation. For each simulated particle, we record its initial momentum $\vec{p}_0$, energy $E_0$, and particle identification $(A_0, Z_0)$, as well as its final energy $E_f$, particle identification $(A_f, Z_f)$, and detection position $\vec{r}_{det}$, which is the point where the particle intersects the surface of the detection sphere. The probability (and consequently the energy spectrum) of detecting a given \ac{CR} emitted within a jet arriving at Earth with $E_f$ depends on the angle $\theta$ between the jet direction and Earth-source line of sight. We account for this considering that each detection position $\vec{r}_{det}$ emulates the Earth position. Therefore, for each detected event, we can compute the angle $\phi$ between the line of sight from the source to the Earth and the particle's emission direction, which is given by
\begin{equation} \label{eq.delta}
    \phi = \arccos\left( \dfrac{\vec{p}_0 \cdot \vec{r}_\text{det}}{\left|\vec{p}_0 \right| \, \left|\vec{r}_\text{det} \right|} \right).
\end{equation}

Finally, events satisfying $\theta - \Psi/2 \leq \phi \leq \theta + \Psi/2$ are considered to represent particles arriving at Earth from a jet with an opening angle $\Psi$ oriented at an angle $\theta$ relative to the Earth-to-source line of sight\footnote{\change{Notice that the observed energy spectrum for $\theta \neq 0^\circ$ must be corrected by a geometric factor, given by the ratio between the cone solid angle and the ring solid angle.}}.

\section{Results}\label{sec:results}

In this section we present the simulation results. To structure the discussion, we fix a few reference parameters, which we later vary to properly discuss their impact. The reference spectral index is $\gamma = 1$, and the injection spectrum given by Eq.~\ref{eq:specInjection} is assumed to be a Heaviside function of the (initial) energy, being 1 below $1\; \text{ZeV}$ and $0$ above.

The combined effect of the emission geometry and magnetic field is quantified through the parameter
\begin{equation}
    \xi \equiv \dfrac{J(E)}{J_0(E_0)} \,,
    \label{eq:xi}
\end{equation}
where $J_0(E_0)$ denotes the injection spectrum, and $J(E)$ refers to the observed spectrum evaluated at an energy $E$, corresponding to a given initial energy $E_0$.

We consider several combinations of the following parameters:
\begin{itemize}[noitemsep, nolistsep]
	\item magnetic field strength ($B$): $10^{-8} \; \text{G}$, $10^{-9} \; \text{G}$, $10^{-10} \; \text{G}$;
    \item coherence length ($L_B$): $100 \; \text{kpc}$, $1 \; \text{Mpc}$, $10 \; \text{Mpc}$;
    \item observation angle ($\theta$): $0^\circ$, $40^\circ$, $80^\circ$.
\end{itemize}
The default value for the jet opening angle is $\Psi = 15^\circ$.

In Sec.~\ref{ssec:res::proton} we present the results for our benchmark scenario, assuming proton primaries. This is followed by Sec.~\ref{ssec:res::nitrogen}, where we present results for nitrogen nuclei. In Sec.~\ref{ssec:res::specIndex} and ~\ref{ssec:res::maxRigidity}, we discuss the dependence of the results on the spectral index and on the maximal magnetic rigidity\footnote{\change{The magnetic rigidity $R$ of a charged particle is defined as $R = E/Ze$, where $E$ is the particle’s energy, $Z$ is the charge number, and $e$ is the elementary charge.}} of the sources, respectively. Section~\ref{ssec:res::isotropy} directly addresses the emission geometry, comparing the isotropic and jetted cases.

\subsection{Pure-proton scenario}\label{ssec:res::proton}

\begin{figure*}[htb]
    \centering
    \includegraphics[width=\linewidth]{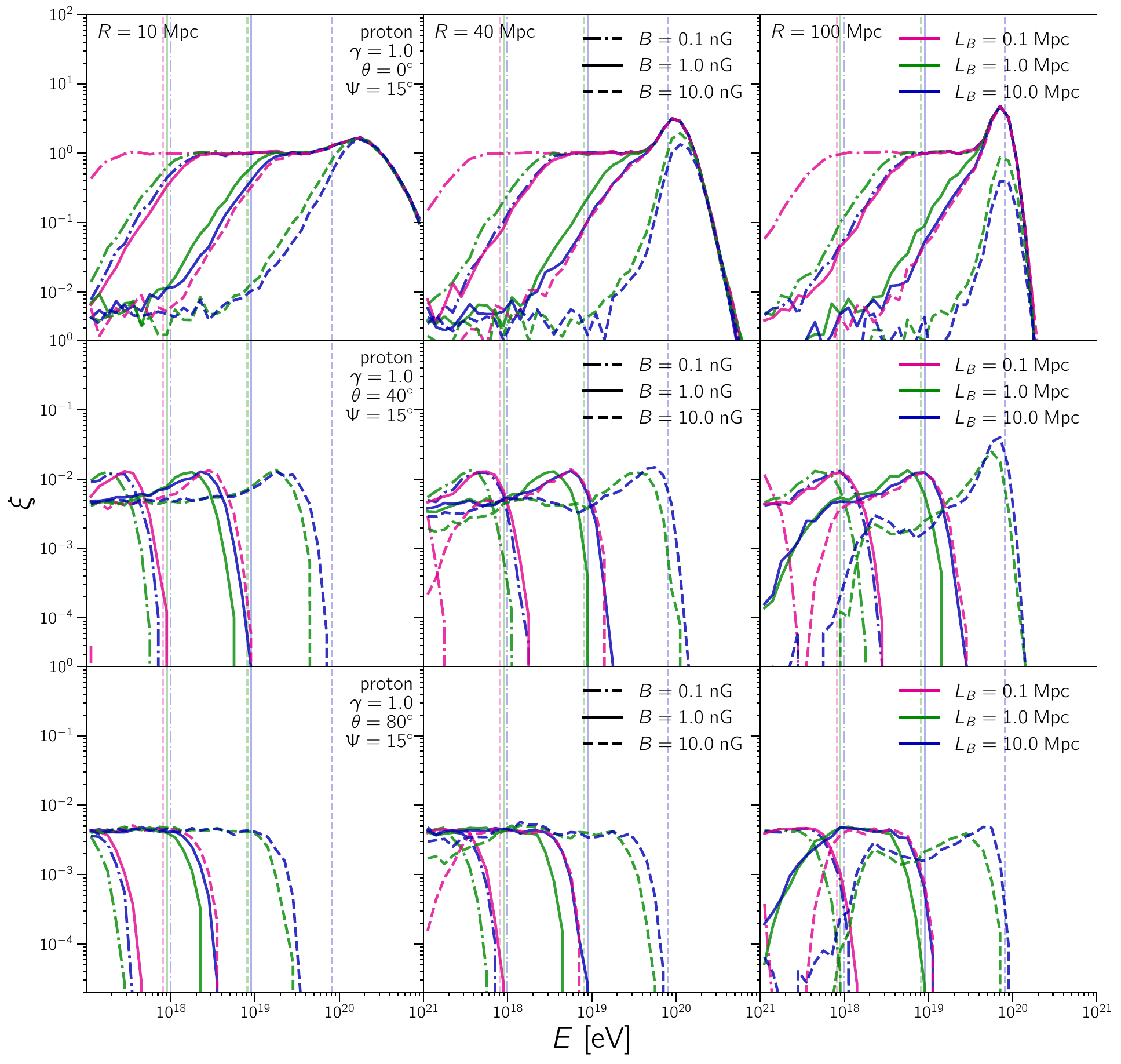}
    \caption{Modification factor obtained for proton primaries. Each column corresponds to a different source-Earth distance ($R$), where $R=$~10, 40, and 100~Mpc, respectively, from left to right. The rows correspond to different viewing angles ($\theta$), namely $\theta=0^\circ$, $40^\circ$, and $80^\circ$, from top to bottom. The jet opening angle is assumed to be $\Psi=15^\circ$.
    Line styles indicate different magnetic field strengths ($B$), while line colors correspond to different coherence lengths, $L_B$. Vertical lines mark the critical energy, defined as in Eq.~\ref{critical_energy}, with the color and line style matching the respective coherence length and magnetic field.}
    \label{fig:modfactor_proton}
\end{figure*}

First, we present the results of simulations involving $10^5$ protons. In Fig.~\ref{fig:modfactor_proton}, they are presented in terms of the modification factor $\xi$ (see Eq.~\ref{eq:xi}). The columns correspond to different values of $R$: the first column represents $R = 10 \; \text{Mpc}$, the second column corresponds to $R = 40 \; \text{Mpc}$, and the third column corresponds to $R = 100 \; \text{Mpc}$. Similarly, the rows indicate different values of $\theta$: the first row represents $\theta = 0^\circ$, the second corresponds to $\theta = 40^\circ$, and the third corresponds to $\theta = 80^\circ$.  The different line styles represent cases where the magnetic field strength is $B = 0.1 \; \text{nG}$ (dot-dashed line), $B = 1.0 \; \text{nG}$ (continuous line) and  $B = 10.0 \; \text{nG}$ (dashed line). The \change{magenta} color represents the results for a coherence length of $L_B = 0.1 \; \text{Mpc}$, while \change{green} curves show results for $L_B = 1.0 \; \text{Mpc}$, and \change{blue} curves show results for $L_B = 10.0 \; \text{Mpc}$. The critical energy, defined by Eq.~\ref{critical_energy}, is marked by a vertical line, with its color and style corresponding to the specified coherence length and magnetic field strength. In certain panels, they are not visible because their values fall outside the displayed range of the figure. In others, there was a superposition among the lines due to the combination of the parameter values. To fix that and improve clarity, an artificial small displacement of $0.1$ was introduced, the $E_\text{c}$ values when $B= 0.1 \; \text{nG}$ were multiplied by an $0.9$ factor and the $E_\text{c}$ values corresponding to $B= 10.0 \; \text{nG}$ by a $1.1$ factor.

The observed peaks in the figures, with $\xi$ reaching values greater than~1 for $\theta = 0^{\circ}$ (first row), are attributed to pile-up effects caused by energy losses due to photopion interactions. Notably, the peak becomes more pronounced as the source distance increases, which reflects the higher probability of photon-pion interactions at greater distances. 

Concerning the magnetic field of $B= 0.1 \; \text{nG}$ (dot-dashed line), it can also be observed that $\xi$ decreases at lower energies, as some particles undergo magnetic deflections larger than \change{$\Psi/2 = 7.5^\circ$} and do not reach Earth when the jet direction is aligned with Earth. Since the deflection increases with the path length, the energy at which $\xi$ deviates from~1 shifts to higher values as the source-Earth distance ($R$) increases for the same parameter values ($B$ and $L_B$). Concerning the cases where the jet is misaligned with Earth (second and third rows) and the magnetic field strength is $B = 0.1 \; \text{nG}$, $\xi$ approaches zero as the energy increases. This behavior occurs because the magnetic field strength is insufficient to deflect these particles coming from jet angles of $\theta = 40^{\circ}$ or $\theta = 80^{\circ}$. However, in these cases, the curves shift to higher energies as the source distance increases, since the angular deflection becomes larger as particles describe longer trajectories. Since the critical energy is proportional to $L_B$, for a given energy and magnetic field strength $B$, particle deflection becomes more pronounced for larger values of $L_B$, as the diffusive regime begins to dominate over the ballistic. This explains why the right side of the curves for the misaligned jets for smaller coherence lengths approaches zero at energies lower than those with higher $L_B$ values. In addition, this is also the reason why the deflection at lower energies is more pronounced for the aligned jet ($\theta = 0^{\circ}$) as $L_B$ increases.
 
The absence of high-energy particles when the jet is not aligned with the Earth direction can be used to place constraints on the relevant parameter space under the hypothesis of a specific single dominant source of \acp{UHECR}. However, this analysis is beyond the scope of the present work and will be addressed in a future work.

Considering a magnetic field strength of $B = 1.0 \; \text{nG}$ (continuous line), the energy at which $\xi$ falls below~1 shifts to higher values for $\theta = 0^{\circ}$ (first row), since magnetic deflection is greater compared to the case with $B = 0.1 \; \text{nG}$ (dot-dashed line). In other words, only the most energetic particles can remain deflected by less than $\Psi/2$ during propagation and still reach Earth. On the other hand, for $\theta = 40^{\circ}$ and $\theta = 80^{\circ}$ (second and third row), the endpoints of the corresponding curves also shift to higher energies, as more energetic particles are able to arrive at Earth, despite experiencing stronger magnetic deflection under $B = 1.0 \; \text{nG}$.

The same effects discussed for magnetic field magnitudes of $B = 0.1 \; \text{nG}$ and $B = 1 \; \text{nG}$ are also present and amplified in the $B = 10 \; \text{nG}$ case. For $\theta = 0^{\circ}$, the energy at which $\xi$ falls below~1 is shifted to higher energies, while the right side of the curves corresponding to $\theta = 40^{\circ}$ and $\theta = 80^{\circ}$ are also shifted to the right. \change{At low energies, the distribution of particle arrival directions becomes isotropic due to magnetic diffusion, and in this limit the modification factors converge to the ratio between the solid angle of a cone with aperture $\psi$ and that of a full sphere $\sim 4 \times 10^{-3}$.} However, for a source distance of $R = 100 \; \text{Mpc}$, the number of protons in the lower energy range decreases and $\xi$ approaches zero as a consequence of the magnetic horizon. This is explained by the long propagation distance of these events in such a highly diffusive regime, resulting in a propagation time longer than the age of the Universe. 

\begin{figure}[htb]
    \centering
    \subfigure[]{
        \includegraphics[width=0.48\textwidth]{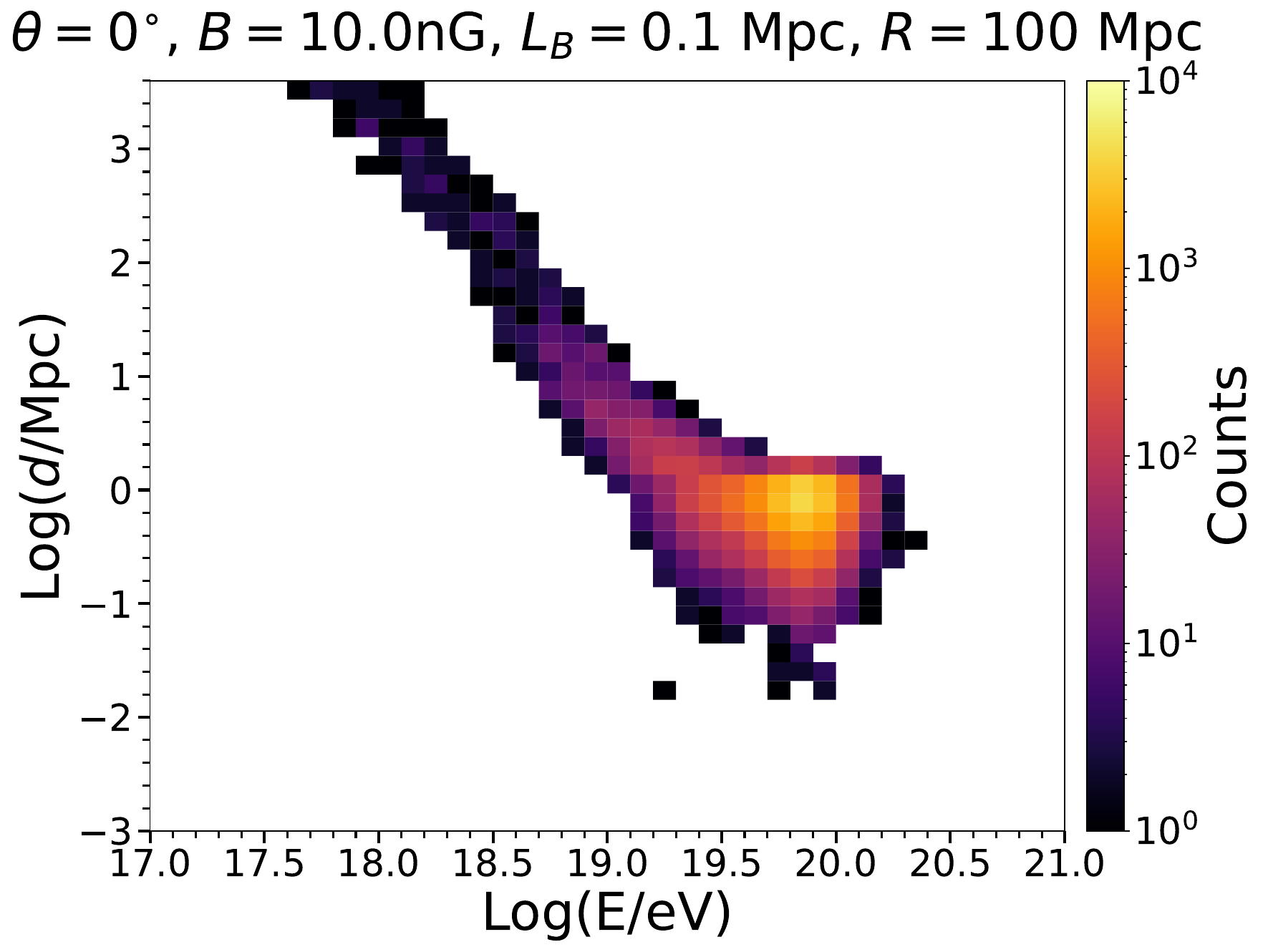}
        \label{fig:elongation_ang0_R100_lc0.1_B10}
    }
    \hfill
    \subfigure[]{
        \includegraphics[width=0.48\textwidth]{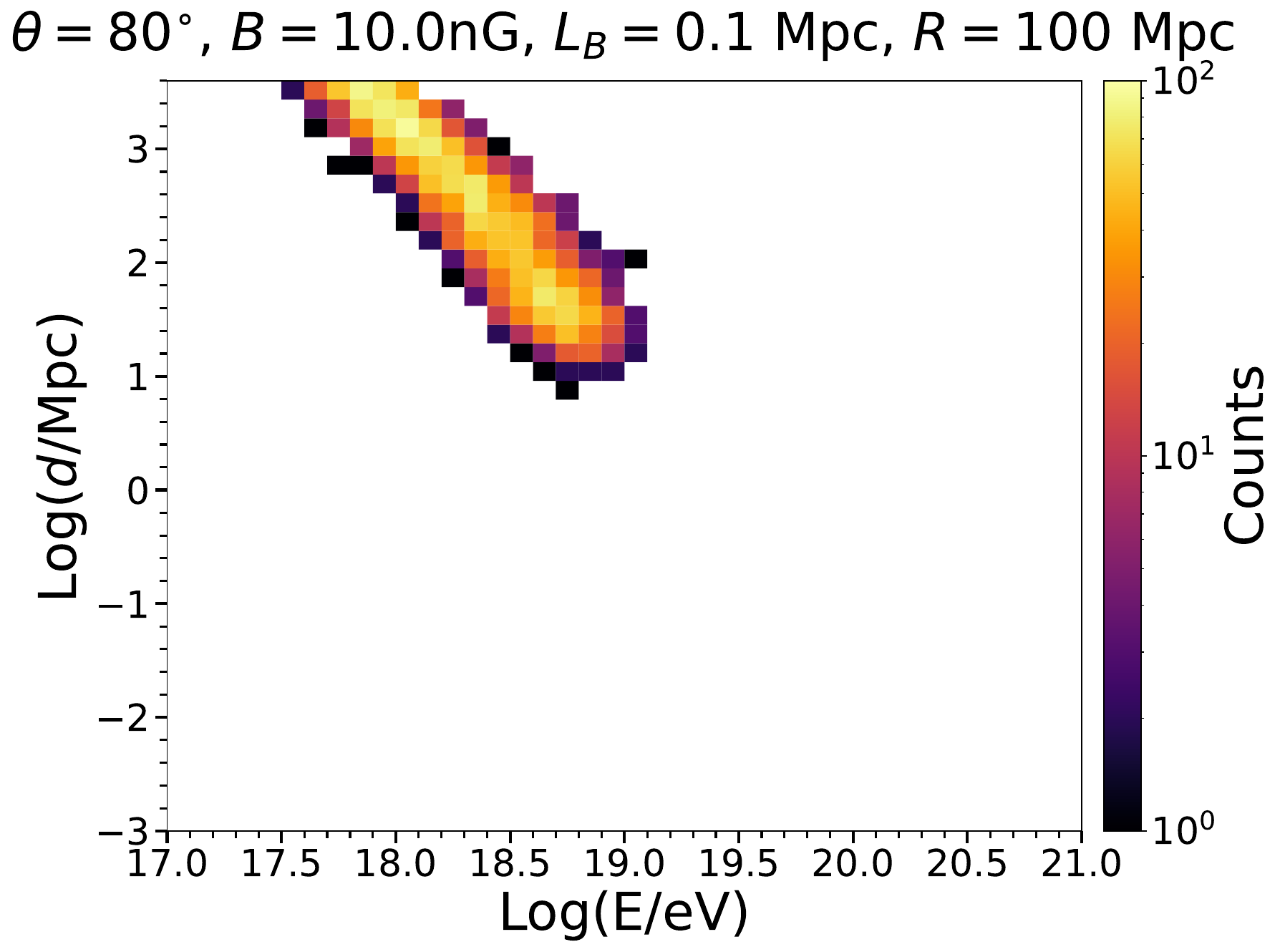}
        \label{fig:elongation_ang80_R100_lc0.1_B10}
    }
    \caption{\justifying Event count distribution as a function of the energy ($E$) and the trajectory elongation ($d$). Figure (a) presents the case where $\theta= 0^{\circ}$, $B= 10.0 \; \text{nG}$, $L_B= 0.1 \; \text{Mpc}$ and $R= 100 \; \text{Mpc}$, while figure (b) shows the results for $\theta= 80^{\circ}$, $B= 10.0 \; \text{nG}$, $L_B= 0.1 \; \text{Mpc}$ and $R= 100 \; \text{Mpc}$.}
    \label{fig:elongation}
\end{figure}

This can be confirmed by analyzing Fig.~\ref{fig:elongation}, where the distribution of events is plotted as a function of energy and trajectory elongation ($d$). The trajectory elongation is defined here as the difference between the distance traveled by the particle and the source-Earth distance:
\begin{equation}
    d = c\tau - R   \,,
\end{equation}
where $\tau$ is the particle's propagation time. Panel a corresponds to $\theta = 0^{\circ}$, while panel b corresponds to $\theta = 80^{\circ}$, both of which show the results for $B= 10 \; \text{nG}$ and $L_B= 0.1 \; \text{Mpc}$. The analysis of panel (a) reveals that most events reaching Earth have a very small elongation length, close to 1~Mpc. This indicates that the actual distance traveled by these particles is nearly equal to the source-Earth distance, resulting from minimal diffusion. As the arrival energy of these events decreases, the elongation length naturally increases. However, the counts for distances of $d \sim 10^3$~Mpc, which is of the same order of magnitude as the distance light would travel during the age of the Universe, are negligible.  We conclude that the suppression observed in the lower energy part of the spectrum is not due to the high diffusion of these particles, which would cause the simulation time to reach the age of the Universe, but rather due to magnetic deflection. Unlike panel a, panel b, corresponding to $\theta = 80^{\circ}$, indicates that a significantly larger number of protons propagated over distances greater than 1~Gpc (not shown in the plot, as the histogram represents the energy of detected particles). Consequently, their propagation times surpass the age of the Universe, preventing them from being detected on Earth.

\subsection{Case of nuclei}\label{ssec:res::nitrogen}

We consider two cases, following what has been done in Sec.~\ref{ssec:res::proton}. The first corresponds to nitrogen primaries, whereas the second refers to iron nuclei. This essentially covers the main band of \ac{CR} rigidities, to enable a proper assessment of the effect of the emission geometry and extragalactic magnetic fields.

\begin{figure*}[htb]
    \centering
    \includegraphics[width=\linewidth]{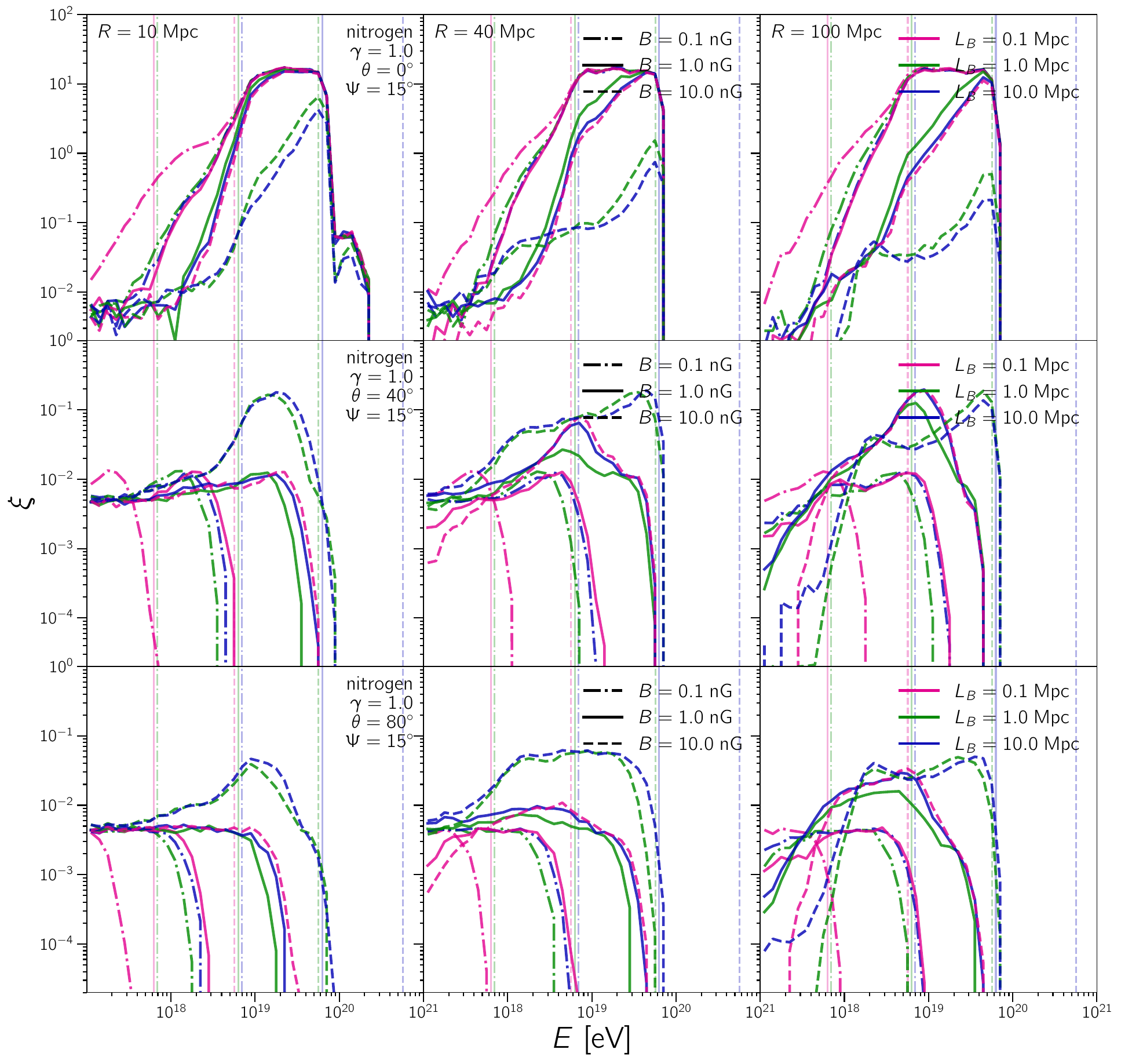}
    \caption{Modification factor obtained for primary \change{nitrogen} nuclei. The columns represent different source-Earth distances ($R$, and the rows vary according to the angle between the jet direction and the line of sight ($\theta$). The line styles indicate different magnetic field strengths ($B$) while the line colors correspond to different coherence lengths ($L_B$). Vertical lines mark the critical energy, with the color and line style matching the respective coherence length and magnetic field.}
    \label{fig:modfactor_nitro}
\end{figure*}

\begin{figure*}[htb]
    \centering
    \includegraphics[width=\linewidth]{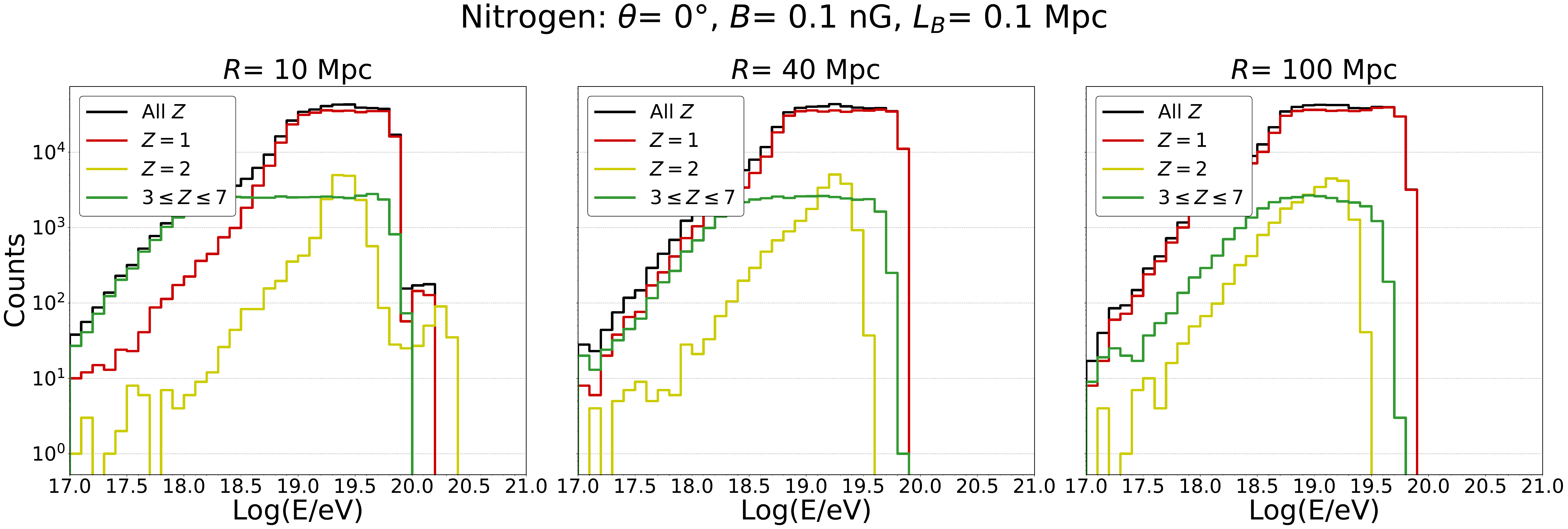}
    \caption{Energy distribution for primary nitrogen nuclei, for different arriving compositions. The figures show the results for the case where the jet direction is $\theta= 0^\circ$, the magnetic strength considered is $B= 0.1$~nG and $L_B= 0.1$~Mpc. The columns vary in source-Earth distance, $R= 10, \; 40, \; 100 \; \text{Mpc}$ from left to right.}
    \label{fig:composition_nitro_B0.1_R_ang0_lb0.1}
\end{figure*}

\begin{figure*}[htb]
    \centering
    \includegraphics[width=\linewidth]{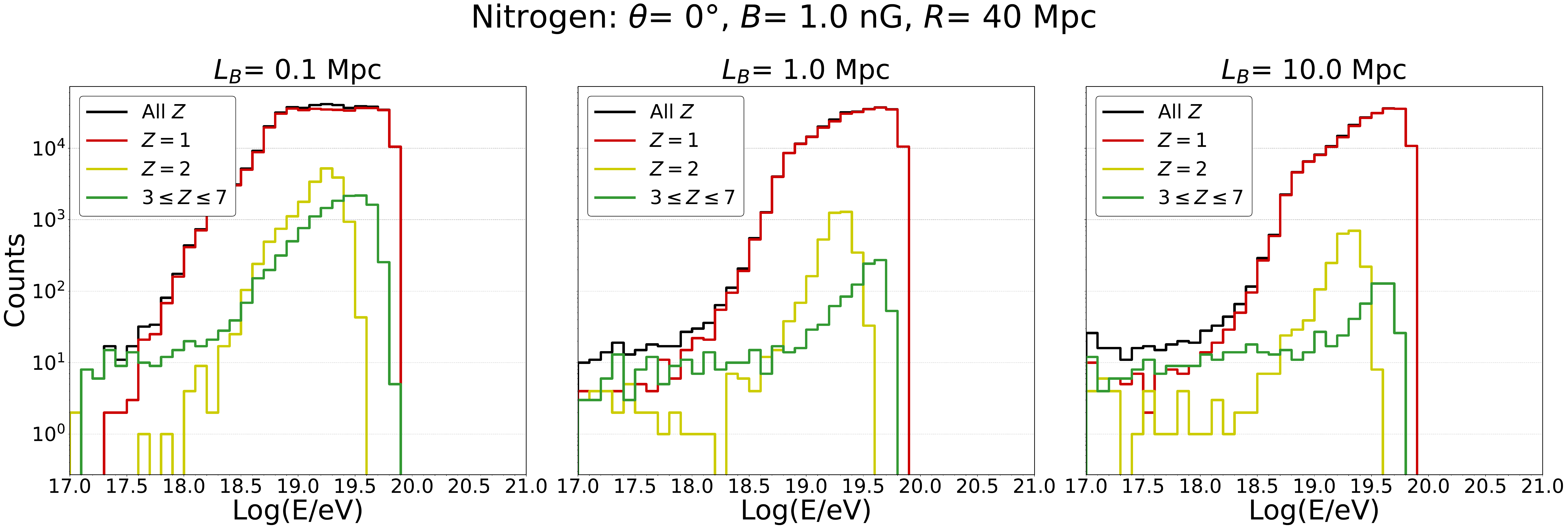}
    \caption{Energy distribution for primary nitrogen nuclei, for different arriving compositions. The figures show the results for the case where the jet direction is $\theta= 0^\circ$, the magnetic strength considered is $B= 1.0$~nG and $R= 40$~Mpc. The columns vary in coherence length $L_B$, $L_B= 0.1, \; 1.0, \; 10.0 \; \text{Mpc}$ from left to right.}
    \label{fig:composition_nitro_B1.0_R_ang0_R}
\end{figure*}

Figure~\ref{fig:modfactor_nitro} shows the modification factor obtained \change{after simulating
$10^5$ nitrogen nuclei}. The panels are organized in the same way as it was for protons, with rows corresponding to angles between the jet direction and the line of sight source-Earth and columns to different source-Earth distances. Due to the complexity introduced by nuclear fragmentation, we show energy distributions at Earth separated by particle type, for selected cases. We first analyze the results corresponding to $\theta = 0^\circ$ (first row). The peak at the highest energies is significantly broader than in the case of proton simulations. This broadening arises because nitrogen nuclei fragment into multiple lighter elements with varying energies, producing a wider spread in the energy distribution of secondary protons arriving at Earth. Figure~\ref{fig:composition_nitro_B0.1_R_ang0_lb0.1} shows the energy spectra for different nuclei, corresponding to the \change{magenta} dot-dashed lines in the first row of Fig.~\ref{fig:modfactor_nitro}. These simulations were performed with an angle between the jet direction and the line of sight of $\theta = 0^\circ$, magnetic field strength $B = 0.1 \; \text{nG}$, and coherence length $L_B = 0.1 \; \text{Mpc}$. From this figure, we observe that for a source distance of $R = 10$~Mpc (first column), the energy range above $10^{20}$~eV is dominated by helium and protons resulting from nitrogen fragmentation. These secondary particles do not travel far enough to undergo photopion production or further nuclear fragmentation. This feature is absent in the spectra for $R = 40$~Mpc (second column) and $R = 100$~Mpc (third column), where the increased propagation distance enhances the probability of interactions with background photons. Additionally, the \change{magenta} dot-dashed curves in the first row of Fig.~\ref{fig:modfactor_nitro} show higher values of $\xi$ at $E \lesssim 6 \times 10^{18}$~eV, due to the combination of nitrogen injected at the source with energies below the fragmentation threshold and an insufficient magnetic field strength and/or coherence length to cause larger deflections. 

\begin{figure*}[htb]
    \centering
    \includegraphics[width=0.65\linewidth]{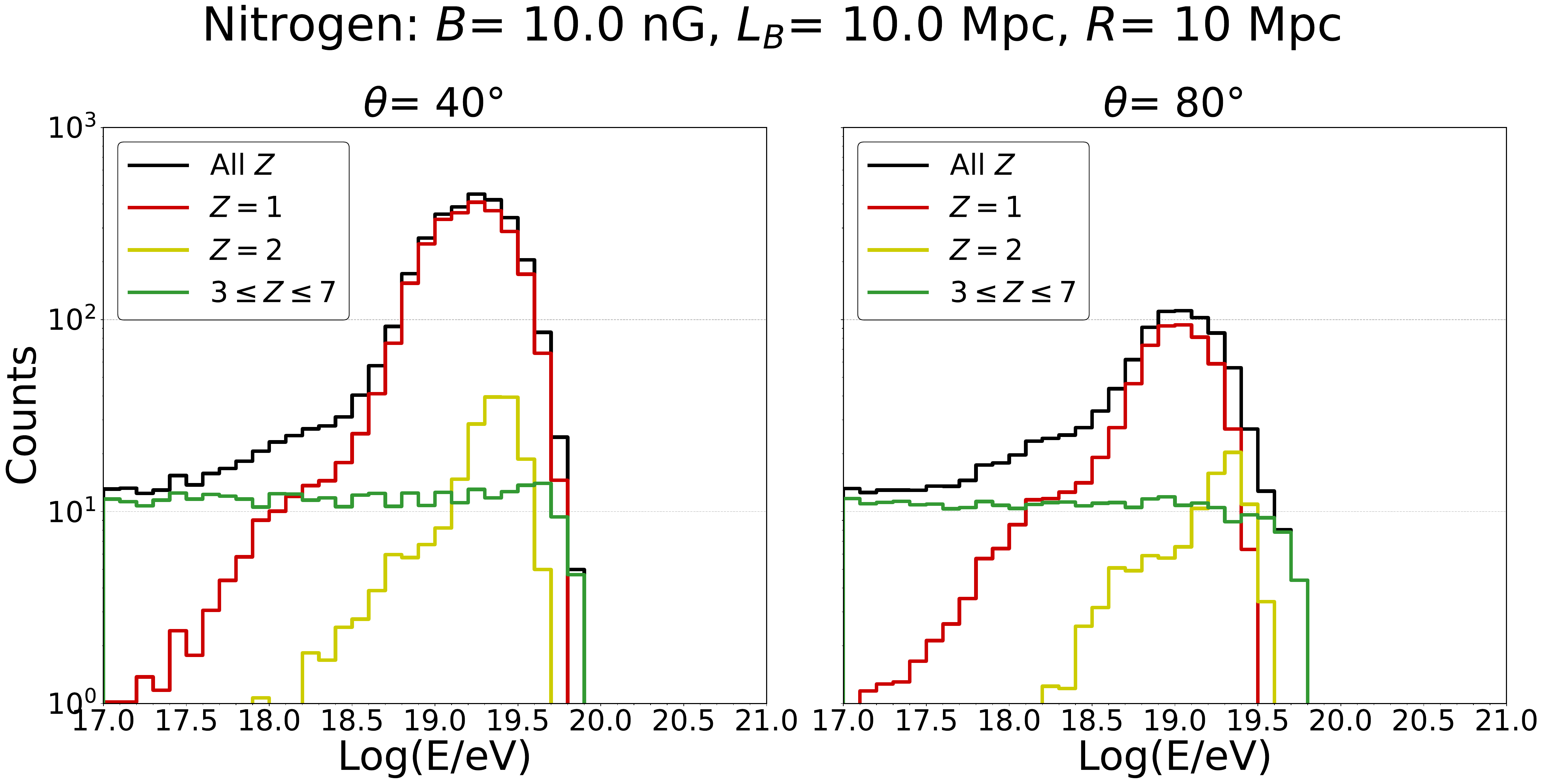}
    \caption{Energy distribution for primary nitrogen nuclei, for different arriving compositions. The panels show the results for the case where the magnetic strength considered is $B= 10.0$~nG, the coherence length $L_B= 10.0 \; \text{Mpc}$, and $R= 10$~Mpc. The left panel corresponds to a jet inclination of $\theta= 40^\circ$ and the right one corresponds to $\theta= 80^\circ$.}
    \label{fig:composition_nitro_B10.0_ang_R10_LB10}
\end{figure*}

When considering other values of $L_B$, the main difference in the energy distribution per nuclear species is that, for higher values of $L_B$, the propagation regime becomes more diffusive at a given energy. As a result, the magnetic deflection enlarges and primary nuclei that would not disintegrate in the $L_B = 0.1 \; \text{Mpc}$ case may now undergo fragmentation as their path length increases. Despite this, the overall conclusions drawn from the analysis remain consistent with those obtained for $L_B = 0.1 \; \text{Mpc}$. As an example case, Fig.~\ref{fig:composition_nitro_B1.0_R_ang0_R} shows a comparison between the energy spectra as a function of the coherence length $L_B$, where $L_B= 0.1, \; 1.0, \; 10.0 \; \text{Mpc}$ from left to right, $B$ is fixed on $1.0 \; \text{nG}$ and $R$ on $40 \; \text{Mpc}$.

\begin{figure*}[htb]
    \centering
    \includegraphics[width=\linewidth]{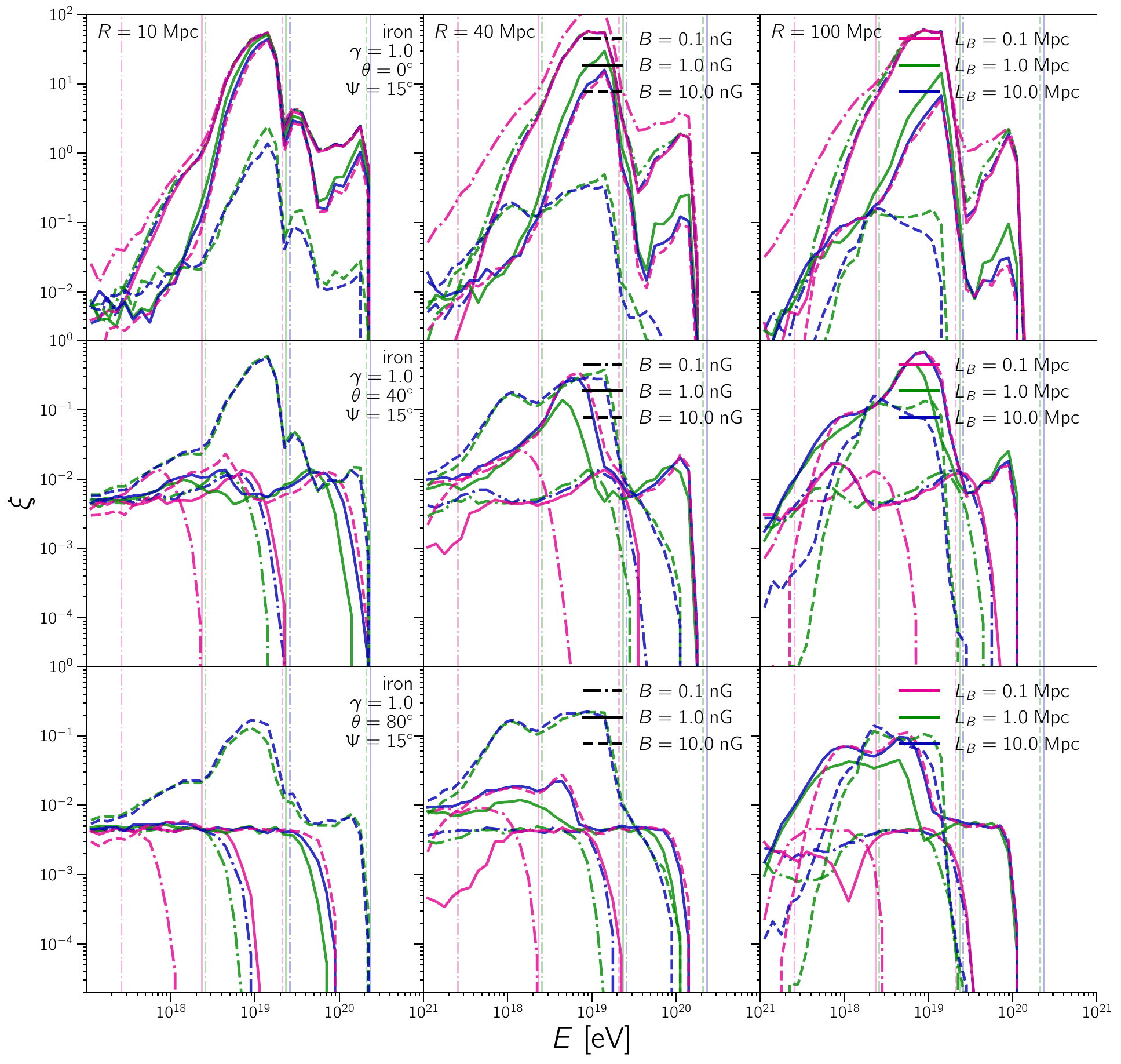}
    \caption{Modification factor obtained for iron. The columns represent different source distances, $R$, and the rows vary in angle between the jet direction and the line of sight, $\theta$. The line styles indicate different magnetic field strengths, $B$, while the line colors correspond to different coherence lengths, $L_B$. Vertical lines mark the critical energy, with the color and line style matching the respective coherence length and magnetic field.}
    \label{fig:modfactor_ferro}
\end{figure*}

With respect to the cases of $\theta = 40^\circ$ (second row), $\theta = 80^\circ$ (third row) and the same magnetic field $B = 0.1 \; \text{nG}$ (dot-dashed line) discussed above,  the curves are dominated by nitrogen emitted with energies below the fragmentation energy threshold. It may be reasonable to assume that the secondary particles produced from nitrogen photodisintegration do not experience enough deflection to reach Earth from misaligned jets with $\theta=40^\circ$ or $\theta=80^\circ$ in the presence of this magnetic field strength. One can also observe that as the source-Earth distance increases, more energetic particles have higher probability of arriving at Earth for the same parameter space due to the large angular deflection. As a consequence, the maximum energy observed in the spectra is shifted to the right, as expected and discussed in the pure-proton scenario. For a similar reason, the increase of $R$ must reveal an increasing contribution of lighter lower-energy nuclei, although it is expected that the primary nuclei remain dominant in the spectra. Similar conclusions are obtained considering $\theta= 80^\circ$.

\begin{figure*}[ht!]
    \centering
    \includegraphics[width=\linewidth]{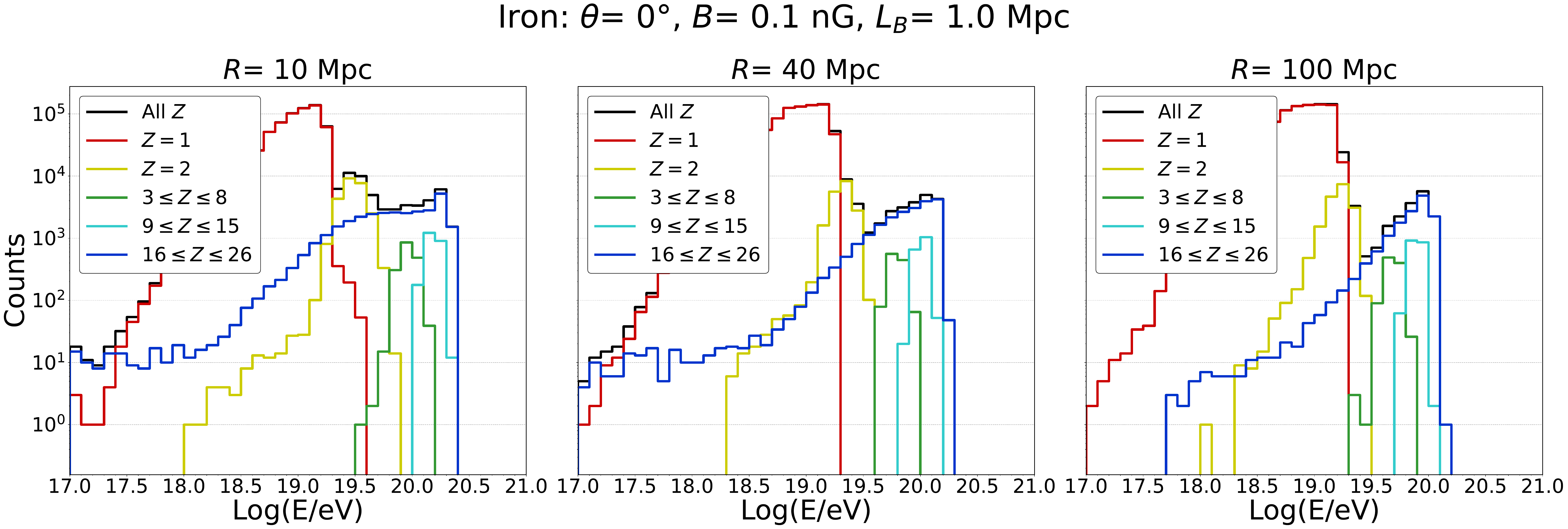}
    \caption{Energy distribution separated by the particle composition for the simulation of iron nuclei as primaries. The figures show the results for the case where the jet direction is $\theta= 0^\circ$, the magnetic strength considered is $B= 0.1$~nG and $L_B= 1.0$~Mpc. The columns vary in source-Earth distance ($R$), $R= 10, \; 40, \; 100 \; \text{Mpc}$ from left to right.}
    \label{fig:composition_ferro_B0.1_R_ang0_lb1.0}
\end{figure*}

For $B = 1 \; \text{nG}$ (continuous line), the peak at the highest energies, observed at $\theta = 0^\circ$ (first row), is narrower for $L_B = 1 \; \text{Mpc}$ and $L_B=10 \; \text{Mpc}$ compared to the peak for $L_B = 0.1 \; \text{Mpc}$ as a consequence of the loss of lower-energy events, resulting from angular deflections greater than $\Psi/2$. For $\theta = 40^\circ$ and $\theta = 80^\circ$, when we compare the results with those corresponding to $B= 0.1 \text{nG}$ (dot-dashed lines), we observe that the maximum energy achieved on Earth also increases. This result is in agreement with the conclusions addressed earlier in the pure-proton scenario. 

The results for the magnetic field strength $B= 10 \; \text{nG}$ show an interesting feature in the modification factor for the misaligned jets. One may see that the curves corresponding to coherence lengths of $L_B = 1 \; \text{Mpc}$ and $L_B = 10 \; \text{Mpc}$ stand out. These higher values of $\xi$ are characterized by secondary protons that are sufficiently deflected to be able to reach Earth. This can be confirmed by Fig.~\ref{fig:composition_nitro_B10.0_ang_R10_LB10}, which shows the energy spectra separated by different arriving compositions for $B= 10.0 \; \text{nG}$, $R= 10 \; \text{Mpc}$, $L_B= 10 \; \text{Mpc}$ and $\theta$ vary in $40^\circ \; \text{(left) and} \; 80^\circ \; \text{(right)}$, as an example case. One may also notice that for $B= 10 \; \text{nG}$, when the distance to the source is $R= 100 \; \text{Mpc}$, the same curves experience a decrease at lower energies. This phenomenon is caused by the magnetic horizon effect, discussed in the proton primary studies. 

Figure~\ref{fig:modfactor_ferro} shows the modification factor resulting after simulating $10^5$ iron nuclei as primary particles. We first analyze the results corresponding to $\theta = 0^\circ$ and $B = 0.1 \; \text{nG}$. The energy spectra for each particle type for these parameters, assuming $L_B= 1.0 \; \text{Mpc}$, are displayed in Fig.~\ref{fig:composition_ferro_B0.1_R_ang0_lb1.0}. For $R = 10 \; \text{Mpc}$, the peak at $E \sim 10^{19} \; \text{eV}$ observed in Fig.~\ref{fig:modfactor_ferro} is dominated by protons as can be seen in Fig~\ref{fig:composition_ferro_B0.1_R_ang0_lb1.0}. In contrast, the peaks at $E \sim 10^{19.5} \; \text{eV}$ and $E \gtrsim 10^{20} \; \text{eV}$ are respectively dominated by helium and iron primaries. For $R = 40 \; \text{Mpc}$ and $R = 100 \; \text{Mpc}$, the helium peak approaches the proton peak, as a result of the higher probability of interaction. When the coherence length $L_B$ increases, the number of events with lower energy decreases due to the larger angular deflection. As expected, the curves corresponding to $\theta = 40^\circ$ and $\theta = 80^\circ$ have a higher maximum energy for larger distances.

For $B=1 \; \text{nG}$ (solid line), the peaks for $\theta = 0^\circ$ result from the contribution of different primary particle families; just as in the previous scenario. Although protons dominate at lower energies, iron and intermediate nuclei prevail at higher energies. Secondary protons cannot reach Earth with energy above $\sim 6 \times 10^{19} \; \text{eV}$ due to the large probability of photopion interaction. The tendency for different values of $L_b$ and $B$ is the same as that seen in the previous analysis of the primary studies of proton and nitrogen (Figs.~\ref{fig:modfactor_proton} and~\ref{fig:modfactor_nitro}\change{)}.

The results for the magnetic field strength $B= 10 \; \text{nG}$ show an interesting feature in the modification factor for the misaligned jets. One may see that the curves corresponding to coherence lengths of $L_B = 1 \; \text{Mpc}$ and $L_B = 10 \; \text{Mpc}$ stand out. These higher values of $\xi$ are characterized by secondary protons that are sufficiently deflected to be able to reach Earth. This behavior is consistent with the nitrogen primary scenarios discussed previously for the same parameter values (Fig.~\ref{fig:modfactor_nitro}). Moreover, the magnetic horizon effect is also evident in this simulation. It is manifested as a suppression of the flux at lower energies when the source-Earth distance is $R = 100 \; \text{Mpc}$.

\subsection{Effect of the spectral index}\label{ssec:res::specIndex}

\begin{figure*}[ht!]
    \centering
    \includegraphics[width=0.32\textwidth]{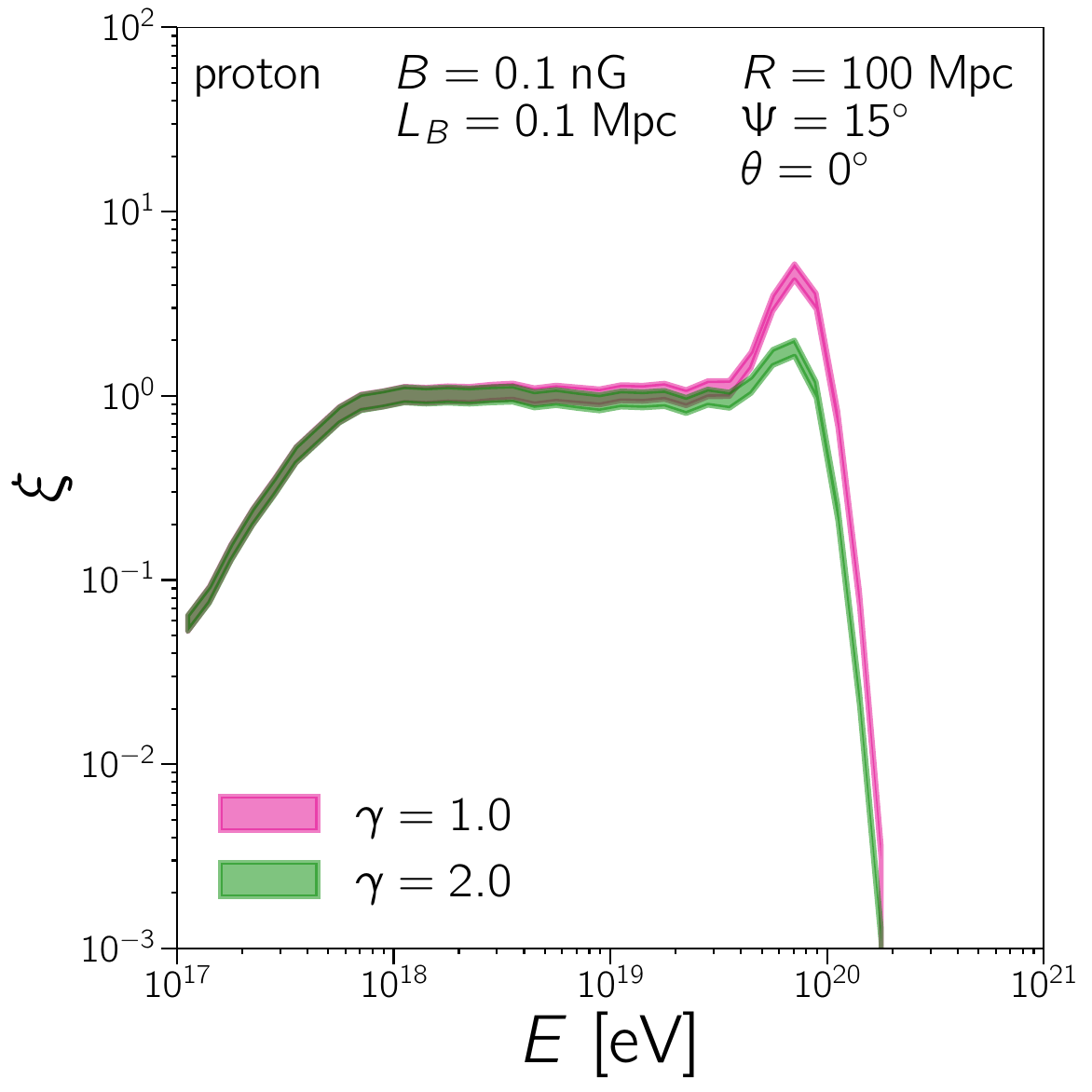}
    \includegraphics[width=0.32\textwidth]{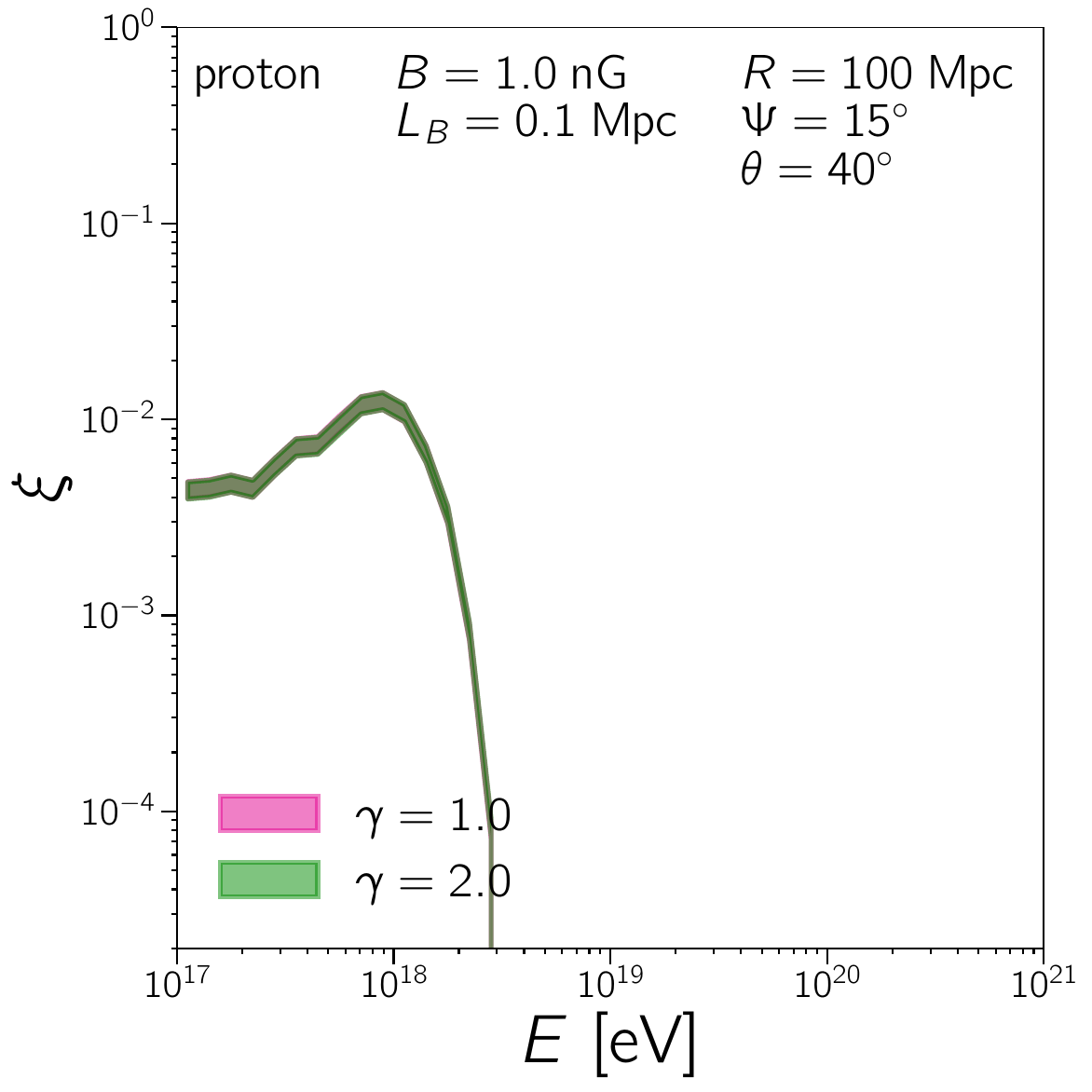}
    \includegraphics[width=0.32\textwidth]{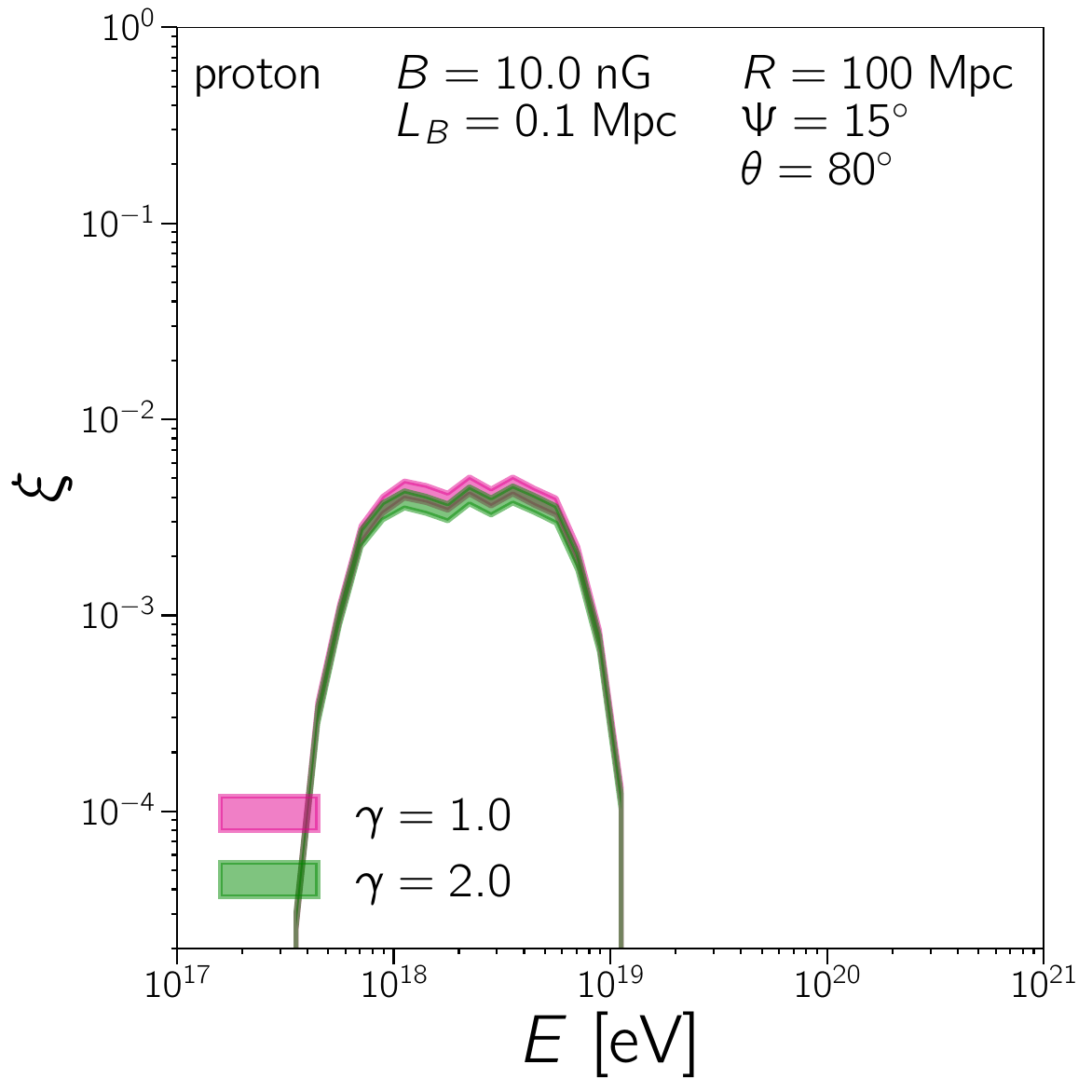}
    \caption{Modification factor obtained for primary protons, assuming an injected energy spectrum with spectral index $\gamma = 1$ (\change{magenta} band) and $\gamma = 2$ (\change{green} band). Each curve band has a $0.1$ width band along the vertical axis. All the results displayed correspond to a distance source-Earth ($R$) of $100$~Mpc, a $0.1$~Mpc coherence length ($L_B$) and a jet opening angle ($\Psi$) of $15^\circ$. The results are displayed for different combinations of magnetic field ($B$) and angle between the jet direction and the line of sight ($\theta$), from left to right: ($B$, $\theta$) $=$ ($0.1$~nG, $0^\circ$), ($1$~nG, $40^\circ$), ($10$~nG, $80^\circ$).}
    \label{fig:compararisonGamma}
\end{figure*}

Our benchmark simulations were done for a spectral index $\gamma = 1$. To investigate the dependence of the results on this parameter, we study another case: $\gamma = 2$. We verified that the modification factor $\xi$ changes only slightly with the spectral index. As an illustration, Fig.~\ref{fig:compararisonGamma} shows a comparison between $\xi$ as a function of energy, considering particles injected at sources with spectral indices $\gamma = 1$ (\change{magenta} band) and $\gamma = 2$ (\change{green} band). To improve the clarity of the plot, as the spectra overlap in some regions, the curves are plotted as bands, each with width $0.1$ along the vertical axis. All plots correspond to $L_B = 0.1 \; \text{Mpc}$ and $R = 100 \; \text{Mpc}$. Panel a depicts the case for $B = 0.1 \; \text{nG}$ and $\theta = 0^\circ$, while panel b shows the results for $B = 1 \; \text{nG}$ and $\theta = 40^\circ$. Simulations for $B = 10 \; \text{nG}$ and $\theta = 80^\circ$ are presented in panel c. 
It can be observed that, except for energies around $\sim 60$~EeV, corresponding to the energy threshold for photopion interactions, the behavior of the curves for both spectral indices is essentially the same, although a more pronounced \ac{GZK} bump can be seen for $\gamma = 1$, due to the larger contribution of more energetic particles compared to the case of $\gamma = 2$.

\subsection{Effect of the maximal rigidity}\label{ssec:res::maxRigidity}

\begin{figure*}[ht!]
    \centering
     \includegraphics[width=0.32\textwidth]{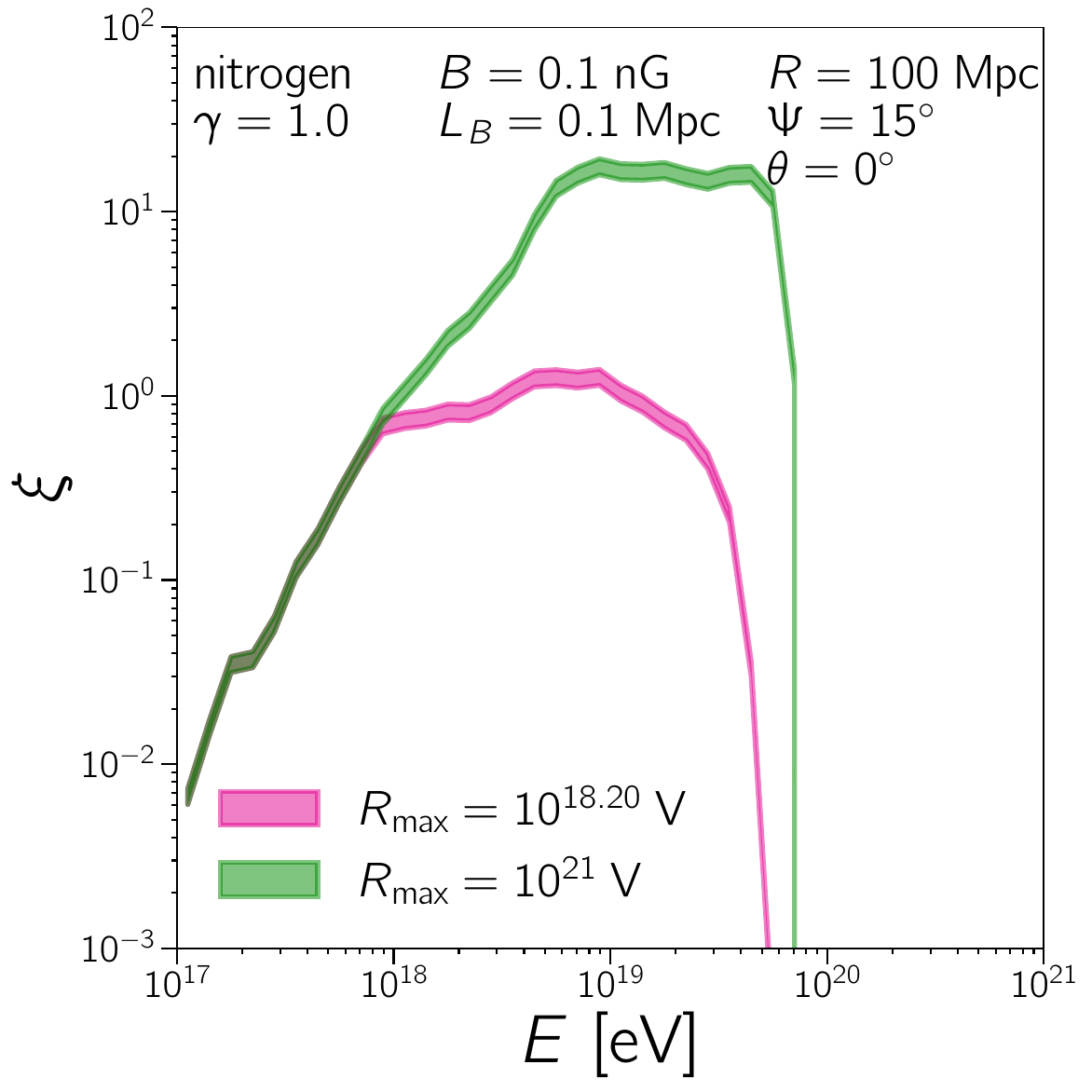}
     \includegraphics[width=0.32\textwidth]{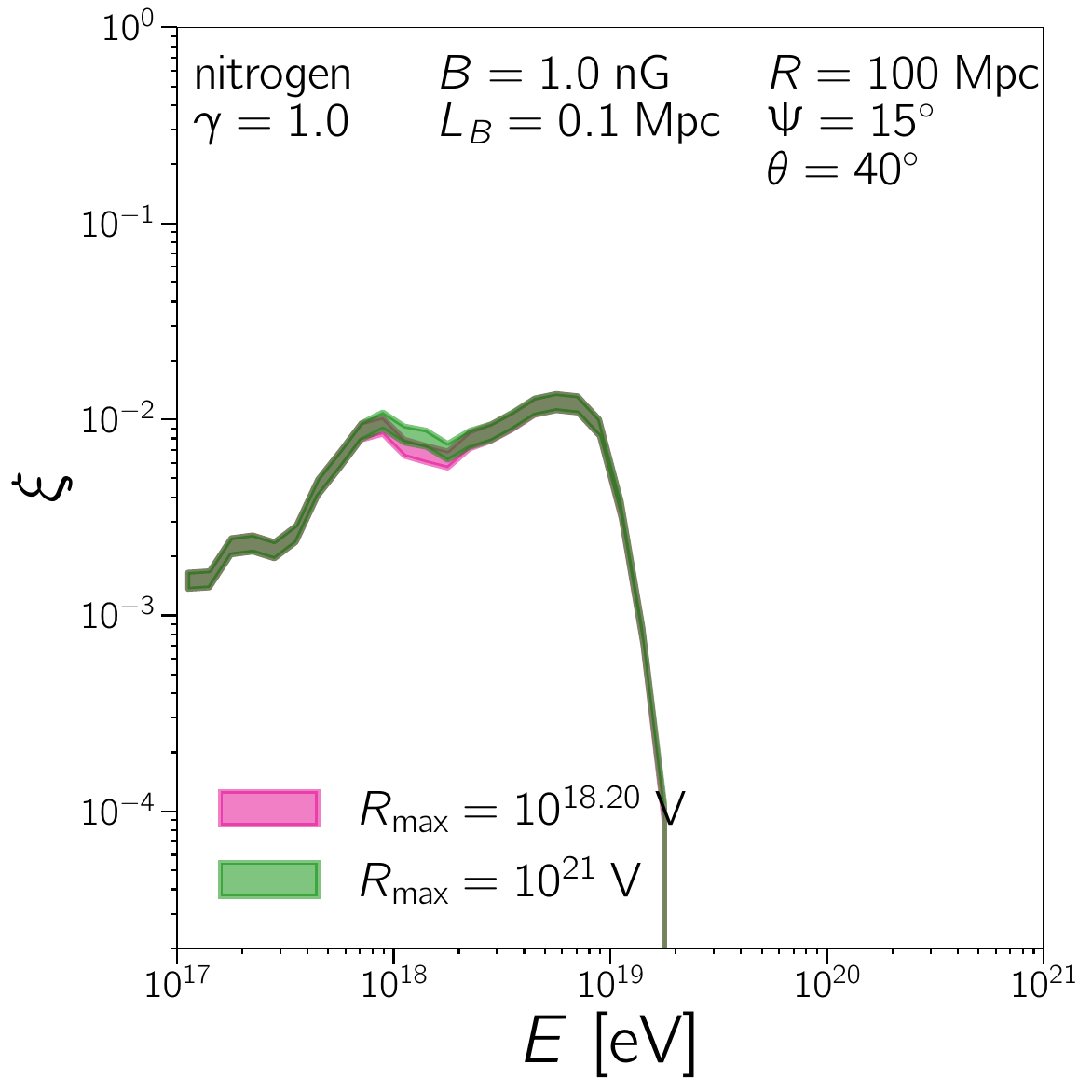}
     \includegraphics[width=0.32\textwidth]{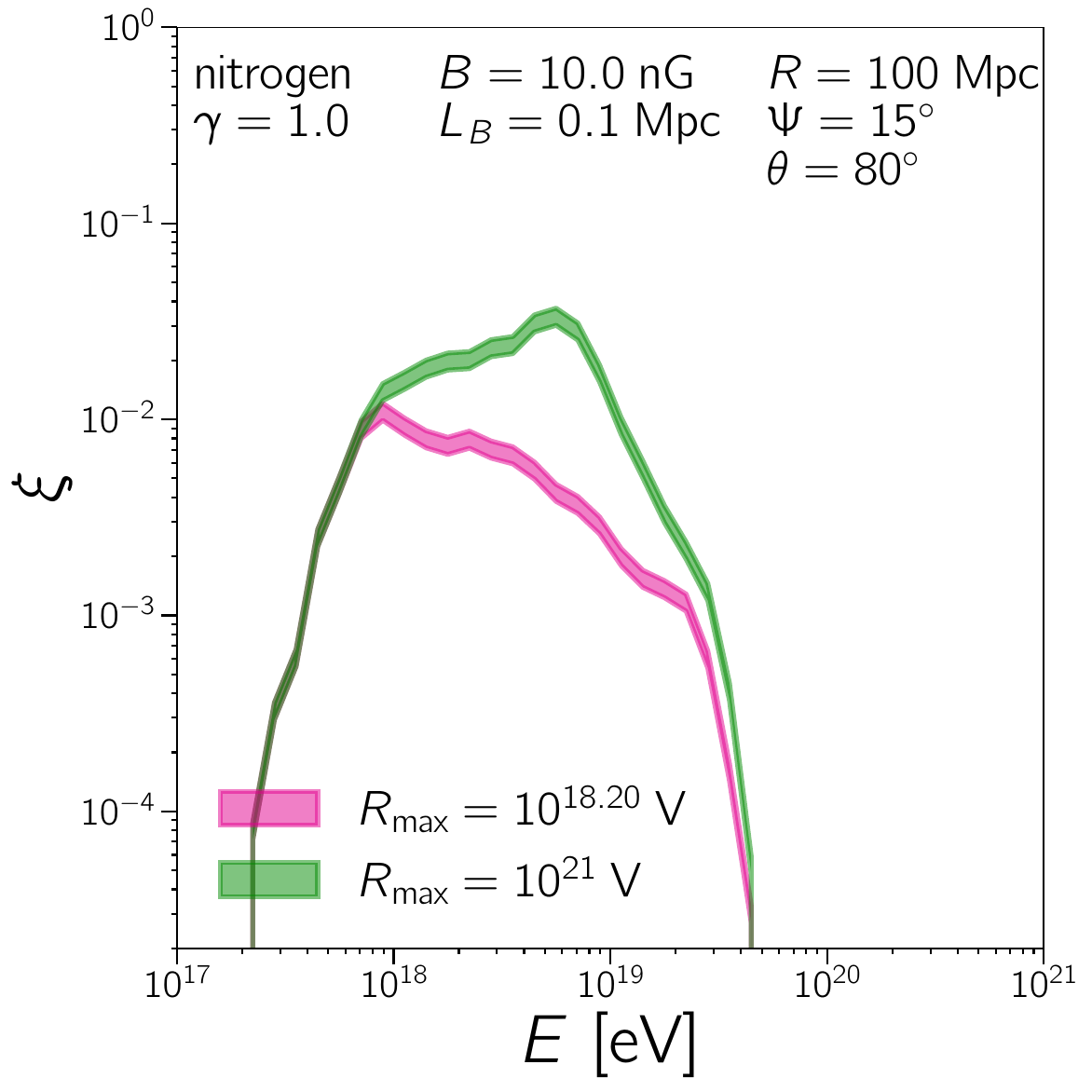}
    \caption{Modification factors for nitrogen nuclei, obtained for a maximal rigidity of $R_\text{max} = 10^{18.2} \; \text{V}$ (\change{magenta} band), and $R_\text{max} = 10^{21} \; \text{V}$ (\change{green band}), for different values of magnetic field ($B$) and various angles between the jet direction and the line of sight ($\theta$). The source is assumed to be distant 100~Mpc from Earth.}
    \label{fig:comparisonRmax}
\end{figure*}

In our benchmark scenario, we neglected the spectral cutoff of the sources. To study this effect separately, we present in Fig.~\ref{fig:comparisonRmax} the modification factors obtained for nitrogen primaries, and for two different maximal magnetic rigidities, $\Rmax = 10^{18.2} \; \change{\text{V}}$ (\change{magenta} band) and $\Rmax = 10^{21} \; \change{\text{V}}$ (\change{green} band), for a fixed spectral index $\gamma=1$. 

\change{The spectral cutoff, according definition in Eq.~\ref{eq:specInjection}, is given by}

\begin{equation}
    {\mchange 
    f_{\text{cut}}(E,Z) = 
    \begin{cases}
        1, & E \leq \Emax \\
        \exp\left(1 - \frac{E}{\Emax}\right), & E > \Emax , 
    \end{cases}}
\end{equation}
\change{with $\Emax = \Rmax \; Ze$.}

There are noticeable differences in the modification factors, for reasons similar to what was discussed in \ref{ssec:res::specIndex}: an interplay between emission geometry, magnetic-field strength, and interactions. The difference is, as expected, more pronounced at the higher energies. For stronger magnetic fields and larger viewing angles (rightmost panel), the difference between the two $\Rmax$ cases studied can be attributed to a change in composition --- primary nitrogen nuclei dominate most of the flux below $E \lesssim 1 \; \text{EeV}$, whereas the contribution of lighter nuclei is only dominant at higher energies.

\subsection{Emission geometry}\label{ssec:res::isotropy}

\begin{figure*}[htb]
    \centering
    \includegraphics[width=\textwidth]{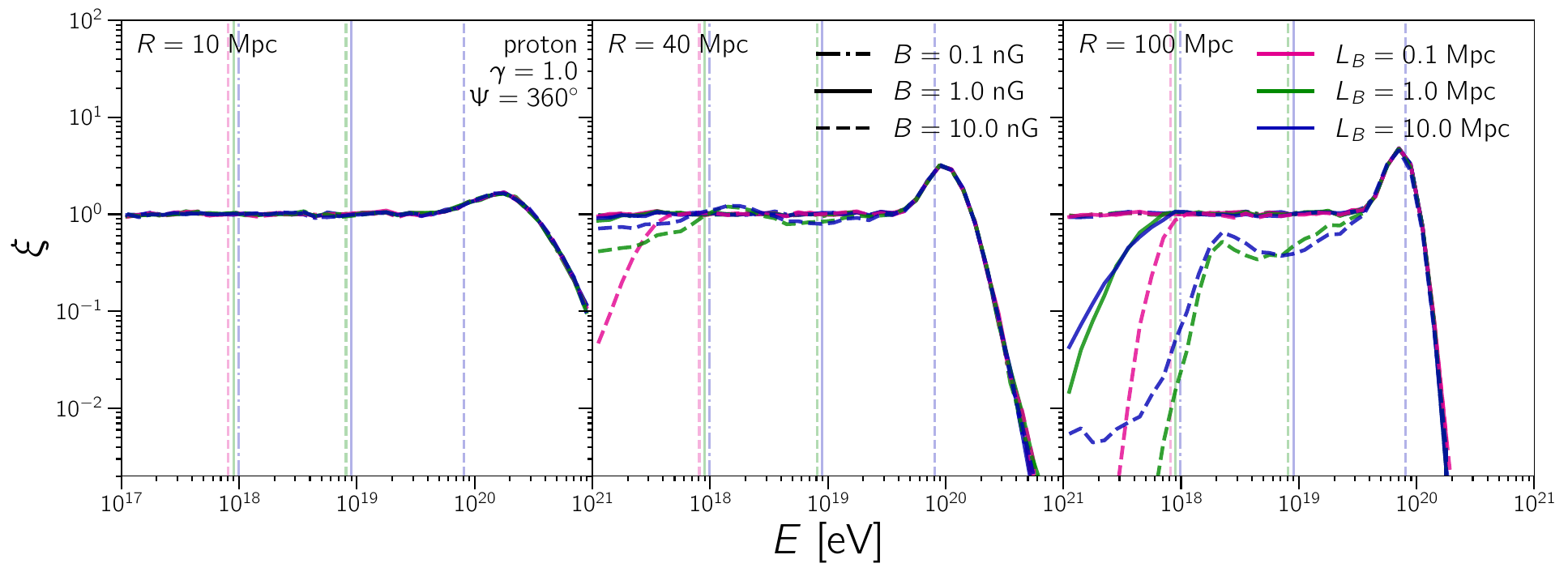}
    \caption{Modification factor for proton primaries assuming isotropic emission. Each column correspond to a different distance source-Earth ($R$), where $R=$~10, 40, and 100~Mpc, respectively from left to right. Line styles indicate different magnetic field strengths ($B$), while line colors correspond to different coherence lengths ($L_B$). Vertical lines mark the critical energy, defined as in Eq.~\ref{critical_energy}, with the color and line style matching the respective coherence length and magnetic field.}
    \label{fig:comparisonGeometry}
\end{figure*}

We also compare the case of an isotropically emitting source to our benchmark (jetted emission) scenario. This is shown in Fig.~\ref{fig:comparisonGeometry}.

The isotropic case has some of the general features of the jetted scenario, shown in Fig.~\ref{fig:modfactor_proton}. For instance, comparing with the aligned jet case ($\theta=0^\circ$), the spectrum is flatter, extending down to lower energies until it is suppressed by magnetic horizon effects. In contrast, the jetted case is strongly peaked for direct emission, becoming slightly flatter with increasing misalignment.

Therefore, the comparison between the isotropic and jetted cases spectra points to a modification factor ($\xi(E)$) that evolves smoothly with energy, magnetic field, coherence length, and source distance. This reflects the averaging of several uncollimated source distributions. Consequently, isotropic emission seems to vary less with the observer's position compared to the jetted case.

\section{Discussion}\label{sec:discussion}

Our study identifies the key role played by \acp{EGMF} in shaping the \ac{UHECR} spectrum. In particular, if the fields are strong enough ($B \gtrsim 1 \; \text{nG}$) and/or if the coherence length is large ($L_B \gtrsim 10 \; \text{Mpc}$), then misaligned jets can also contribute to the total spectrum. As a consequence, sources that are not seen at extreme energies (above $\sim 50 \; \text{EeV}$) can be important contributors to the lower-energy part of the spectrum.

Some of the scenarios studied here may not be very realistic. For instance, over distances of $\sim 100 \; \text{Mpc}$, the contribution of cosmic voids to the volume filling factors of magnetic fields likely increases~\cite{vazza2017a}, such that some combinations of parameters such as $R \gtrsim 100 \; \text{EeV}$ and $B \gtrsim 10 \; \text{nG}$ are not possible in reality, since upper limits for magnetic fields in voids are $47 \; \text{pG}$~\cite{jedamzik2019a}. 

We assumed a Kolmogorov power-law spectrum for the magnetic field, which may not really be an adequate description of the turbulent intergalactic medium. In fact, cosmological magnetohydrodynamical simulations of \acp{EGMF} suggest that magnetic spectrum can affect \ac{UHECR} propagation~\cite{alvesbatista2017c, hackstein2018a}. Furthermore, we only investigated coherence lengths between 100~kpc and 10~Mpc, compatible with the (weak) existing limits~\cite{alvesbatista2020a}, although this quantity can vary by several orders of magnitude~\cite{alvesbatista2021a, durrer2013a, vachaspati2021a}. 

In addition, our results should be interpreted considering the volume filling factor of magnetic fields in between the sources and Earth. The combined impact of emission geometry and \acp{EGMF} on the \ac{UHECR} spectrum is evidently more pronounced for stronger fields and larger distances. Nevertheless, strong magnetic fields are not volume filling, such that their influence is limited to only a small subset of propagation paths. Moreover, a detailed three-dimensional model of \acp{EGMF} that would allow us to properly model their effect on \acp{UHECR} is lacking.~\cite{coleman2023a}. 
Cosmological \ac{MHD} simulations and Faraday-rotation measurements consistently indicate that regions with $B \gtrsim 10^{-7}\,\text{G}$ occupy at most a few percent of the comoving volume, being concentrated in galaxy clusters and the densest parts of filaments~\cite{osullivan2020a, locatelli2021a, vazza2021b}. As a result, only the spectra of sources embedded in such environments or viewed by an observer embedded therein will be significantly affected~\cite{sigl2004a, armengaud2005a, alvesbatista2017c}. 

Due to energy attenuation, highest energy events ($\gtrsim 60$ EeV) are expected to come from sources within $\sim \text{100}$~$\text{Mpc}$. As a consequence, the single-source scenario studied here is consistent with the estimated density of \acp{UHECR} sources for these events~\cite{auger2013b, bister2024constraints}. However, considering the lower-energy events, the contribution of more distant objects to the energy spectrum is believed to be dominant and this is precisely where the effects of emission geometry and magnetic fields are the strongest. Consequently, one may argue that our single-source study should be less realistic for these events. However, we must keep in mind that these studies on lower limits of source density only considered isotropic emission. As shown in this work, the inclusion of jet emission can be decisive in determining the energy spectrum observed at Earth. Therefore, introducing this geometry should not lead to the same conclusions as those derived under the assumption of isotropy.

A remarkable consequence of our results is that, even for jets pointing toward us, the spectrum is significantly modified at the highest energies ($E \gtrsim 50 \; \text{EeV}$) for coherence lengths comparable to the source distance, as is the case of the upper rows of Figs.~\ref{fig:modfactor_proton}, \ref{fig:modfactor_nitro}, and \ref{fig:modfactor_ferro}. This effect arises from the extended propagation paths and increased interaction probabilities. This is important for nuclei, since it could affect the mass composition measured at Earth. 

\change{In our analysis we averaged over several realizations of the turbulent magnetic field. In reality, if only a single realisation were considered, features such as flux magnification or demagnification could appear~\cite{harari1999a, harari2002a, dolgikh2023a, dolgikh2024a}. Our choice was motivated by computational tractability: sampling the full anisotropic jet emission in combination with multiple \ac{EGMF} realizations is extremely costly. The adopted approach provides a reasonable first-order approximation to the \textit{average} \ac{UHECR} intensity at Earth, but it inevitably suppresses the variance in modification factors that could arise in individual realizations of the field. Such effects are clearly visible in the numerical simulations of Refs.~\cite{dolgikh2023a, dolgikh2024a}, where energy-dependent (de\-)magnification appears in certain directions.}

\change{Causticlike structures and multiple images may be particularly relevant when the source distance is comparable to the coherence length of the magnetic field ($R \sim L_B$). In that regime, one expects sharp, direction-dependent amplifications or suppressions of the flux, as well as spectral distortions, which are averaged out here. Nevertheless, this does not compromise the analysis here presented, which demonstrates the initial hypothesis that the \ac{UHECR} is affected by the interplay of emission geometry and \acp{EGMF}. A dedicated study with fixed magnetic-field realizations and true jet emission would be required to capture these lensing effects in detail.}

\change{We have also neglected the \ac{GMF}, which is known to induce causticlike effects~\cite{harari1999a, harari2002a, prouza2003a, bister2024a} depending on the relative strength of the turbulent component with respect to the regular.}

The effects discussed here are also important for understanding the transition between Galactic and extragalactic cosmic rays. On one hand, magnetic horizons suppress the extragalactic component at lower energies. On the other hand, the interplay of \acp{EGMF}, interactions, and emission geometry may increase the contribution of specific sources at certain energy bands, rendering the interpretation of the spectrum more complex.

The time it takes for \acp{CR} to reach Earth from their sources, compared to their distance, also has important implications for multimessenger observations. For instance, \acp{CR} magnetically trapped in galaxy clusters undergo more interactions with the intracluster gas when magnetic fields are stronger, consequently increasing the flux of secondary particles such as neutrinos and gamma rays~\cite{hussain2021a, hussain2023a, hussain2024a}. The same happens in intergalactic space -- the longer \acp{CR} travel, the more likely they are to produce cosmogenic photons and neutrinos. This could, in principle, completely alter the current estimates of cosmogenic neutrino and photon fluxes~\cite{alvesbatista2019a, heinze2019a}, significantly increasing them.

Phenomenological fits to the \ac{UHECR} data~\cite{auger2017a, alvesbatista2019a, heinze2019a} are commonly done in one dimension, such that the emission geometry is completely neglected. Nevertheless, even in these simple one-dimensional models, one could think of the trajectory elongation and energy losses as a term that effectively changes the \textit{apparent} intrinsic properties of the source like $\Rmax$ and $\alpha$, which then become $\Rmax = \Rmax(B, L_B, D, Z)$ and similarly for the spectral index $\alpha$. 
In fact, this has been recently done in Ref.~\cite{auger2024e}. Nevertheless, the hypothesis of isotropic emission implicitly remained. This also has implications for the fluxes of cosmogenic neutrinos and photons discussed mentioned in the previous paragraph, as discussed in Ref.~\cite{alvesbatista2025b}.

This leads to a degeneracy problem: two sources with very different intrinsic properties could appear similar to an observer, solely due to differences in the intervening magnetic structure and emission geometry. Disentangling such effects requires better knowledge of the local \ac{EGMF}, or additional input from other messengers.

\section{Conclusions}\label{sec:conclusions}

In this work, we investigated the influence of emission geometry and \aclp{EGMF} on the observed energy spectra of \acp{UHECR}, assuming they are emitted in astrophysical jets. Our findings provide novel insights into the conditions under which \ac{UHECR} events from jets can be observed on Earth and reveal how the interplay between emission geometry and magnetic fields critically shapes the observed energy spectra. 

An immediate consequence of our results is a decrease in the energy of the \acp{UHECR} arriving at Earth resulting from the elongation of their trajectories, caused by the intervening magnetic fields. As a consequence, unsurprisingly, there is also a substantial change in the composition of the particles reaching Earth.

We have shown that jetted sources emitting toward Earth might not contribute to the \ac{UHECR} spectrum at specific energy ranges due to a combination of emission geometry, interaction lengths, and magnetic horizons. Off-axis sources whose jets are misaligned with respect to our line of sight can significantly contribute to the observed spectrum, especially at energies $\lesssim 10 \; \text{EeV}$. This challenges the intuitive assumption that aligned sources always dominate the flux. It also poses problems for phenomenological models that rely on the hypothesis of isotropic emission.

As a result, the observed arrival direction and composition of \acp{UHECR} may not reflect the true properties of the emitting sources. Instead, a significant fraction of the flux could originate from sources that leave no counterpart in other messengers, namely photons and neutrinos, complicating efforts to correlate cosmic rays with known astrophysical objects.

This also implies a potential observational bias: some events may under-represent important sources contributing to the bulk of the \ac{UHECR} spectrum at the highest energies, simply because their emission is misaligned and/or attenuated. Thus, interpreting measurements without accounting for geometric effects may lead to an incomplete view of source candidates. This is particularly critical for composition-sensitive observables, where magnetic deflections and spectral distortions can mimic or mask source characteristics.

In this work, we have not investigated the arrival directions of \acp{UHECR}. Nevertheless, the drastic spectral changes we found strongly suggest that the anisotropies would follow suit, especially at the highest energies. This will be deferred to future work.

As an example of the potential applications of this study, we demonstrated that, under the hypothesis of a single dominant source of \acp{UHECR}, for a magnetic field strength of $B = 0.1 \; \text{nG}$, the absence of high-energy particles when the jet is not aligned with the Earth's direction can be used to place constraints on key parameters such as source distance, coherence length, and magnetic field strength. These results not only enhance our understanding of \ac{UHECR} propagation but also provide a framework for future studies aiming to identify and characterizing potential \ac{UHECR} sources.

Future studies should aim to incorporate realistic three-dimensional \ac{EGMF} models derived from constrained cosmological simulations, ideally combined with considerations about the emission geometry directly at the source level. However, this is manifestly difficult, and this type of information is lacking. Bridging these aspects will be crucial to reduce degeneracies and improve our ability to reconstruct the source properties of \acp{UHECR}. 

\section*{Data Availability}

The data that support the findings of this article are not publicly available upon publication beacause it i not technically feasible and/or the cost of preparing, depositing, and hosting the data would be prohibitive within the terms of this research project. The data are available from the authors upon reasonable request.

\section*{Acknowledgments}

The work  of S. S. S., R. M. d. A. and J. R. T. d. M. N. is supported by the Conselho Nacional de Desenvolvimento Científico e Tecnológico - CNPq Grant No. 406825/2023-8 and Fundação Carlos Chagas Filho de Amparo à Pesquisa do Estado do Rio de Janeiro - FAPERJ Grant No. E-26/210.091/2023. R. A. B. is supported by the Agence Nationale de la Recherche (ANR),
Project No. ANR-23-CPJ1-0103-01.

\newpage

\label{lastpage}
\end{document}